\documentclass[12pt,preprint]{aastex}

\usepackage{graphicx}
\usepackage{epstopdf}

\newcommand{\kms}{km s$^{-1}$}
\newcommand{\ha}{H$\alpha$}
\newcommand{\solar}{\ifmmode_{\sun}\else$_{\sun}$\fi}

\newcommand{\HII}{H$\,${\sc ii}}
\newcommand{\HI}{H$\,${\sc i}}
\newcommand{\rd}{$R_{\rm D}$}
\newcommand{\rbr}{$R_{\rm Br}$}
\newcommand{\rfuv}{$R_{\rm FUV}$}
\newcommand{\rha}{$R_{\rm H\alpha}$}
\newcommand{\rcl}{$R_{\rm cl}$}

\newcommand{\alphavir}{$\alpha_{\rm vir}$}

\begin{document}

\title{\HI\ Clouds in LITTLE THINGS Dwarf Irregular Galaxies}

\author{Deidre A. Hunter}
\affil{Lowell Observatory, 1400 West Mars Hill Road, Flagstaff, Arizona USA}

\author{Bruce G.\ Elmegreen}
\affil{IBM T.\ J.\ Watson Research Center, PO Box 218, Yorktown Heights, New York}

\and

\author{Clara L.\ Berger\altaffilmark{1}}
\affil{Lowell Observatory, 1400 West Mars Hill Road, Flagstaff, Arizona USA and Northern Arizona University, Flagstaff, Arizona}

\altaffiltext{1}{Current address: Wellesley College, Wellesley, Massachusetts}

\begin{abstract}
We identify 814 discrete \HI\ clouds in 40 dwarf irregular galaxies from the
LITTLE THINGS survey using an automated cloud-finding algorithm.
The cloud masses range from $\sim10^3$ to
$10^7\;M_\odot$, have a surface density averaged over all of the clouds of $\sim9.65\;M_\odot$
pc$^{-2}$, and constitute 2-53\% of the total \HI\ mass of the host galaxy.
For individual clouds, the mass including He varies with cloud
radius as $\log M_{\rm gas} = (2.11\pm0.04)\times \log R_{\rm cl} +
(0.78\pm0.08)$ and the internal velocity dispersion varies as $\log V_{\rm
disp} = 0.5 \times \log R_{\rm cl} - 0.57\pm0.21$. The \HI\ clouds tend to be in the
outer regions of the galaxies, with 72\% of the galaxies having more than 70\%
of their clouds outside one disk scale length, and 32\% of the galaxies having
more than 50\% of their clouds outside the radius encircling the  \HII\ emission.
36\% of the clouds are essentially non-self-gravitating from \HI\
alone, with a virial parameter that exceeds $\alpha_{vir}\sim10$,
and 5\% have $\alpha_{vir}\le 2$.
We estimate the missing molecular mass, based on the total star formation rate and a
typical molecular consumption time of 2 Gyr, as observed in CO-rich galaxies.
The resulting molecular fraction has a value averaged over the galaxies of 0.23 and correlates
with both the surface density of star formation and the fraction of \HI\ clouds
in the outer regions. We conclude that a significant fraction of the inner
parts of these dwarf galaxy disks is in the form of dark molecular gas, and
that this fraction could be high enough to make the inner disks mildly
gravitationally unstable as a precursor to star formation.
\end{abstract}


\section{Introduction} \label{sec-intro}

It is commonly believed that stars form in molecular clouds and molecular clouds
form from \HI\ clouds \citep{mckee07}.
This concept is difficult to observe in gas-rich dwarf irregular galaxies (dIrrs)
because molecular tracers like CO and cold dust are usually too faint to see at
the low metallicities of these galaxies
\citep[e.g., $\le15$\% solar; for example,][]{wlmco1,cormier17,review19}.
Generally only \HI\ is observed in emission and that \HI\ has a large-scale average gas
density that is too low for clouds to form by self-gravitational processes, which
seem to be more evident in spirals \citep[see, for
example,][]{vanzee97,aomawa98,bigiel08}. Yet, dIrrs have both star formation and
\HI\ clouds. For example, \citet{lmc} conducted a survey of neutral hydrogen
clouds in the Large Magellenic Cloud (LMC) and identified 468 clouds with masses
in the range of $10^3$ $M$\solar\ to $10^5$ M\solar.

Here, we present a study of the \HI\ clouds in a sample of 40 nearby dwarf
irregular galaxies. The clouds are found from radio interferometric
position-velocity \HI\ data cubes using automatic cloud detection software made
publicly available by \citet{cprops}. We discuss the source of the data in
Section \ref{sec-data}, and the cloud identification process in Section
\ref{sec-id}. In Section \ref{sec-results} we present the cloud properties and
distributions within the galaxies. Although we compare our results to those of
\citet{lmc}, the galaxies in our sample had spatial resolutions ranging from of
19 pc (IC 10) to 424 pc (F564-V3) while Kim et al.'s survey had a spatial
resolution of 12 pc at the LMC. Section \ref{dark} then considers the likelihood
that a significant fraction of the gas is molecular in the inner regions.

\section{Data} \label{sec-data}

We used the LITTLE THINGS\footnote[2]{Funded in part by the National Science Foundation through grants AST-0707563, AST-0707426, AST-
0707468, and AST-0707835 to US-based LITTLE THINGS team members and with generous technical and
logistical support from the National Radio Astronomy Observatory.}
(Local Irregulars That Trace Luminosity
Extremes, The \HI\ Nearby Galaxy Survey; Hunter et al. 2012) sample of dIrr galaxies.
LITTLE THINGS is
a multi-wavelength survey of 37 dIrr galaxies and 4 Blue Compact Dwarfs (BCD) aimed at
understanding what drives star formation in tiny systems. The LITTLE THINGS galaxies
were chosen to be nearby ($\le$10.3 Mpc), contain gas so they could be forming stars, and cover
a large range in dwarf galactic properties, such as rate of star formation.
The LITTLE THINGS data sets include \HI-line maps obtained with the National Science Foundation's Karl G.\ Jansky Very Large Array
(VLA\footnote[3]{The VLA is a facility of the National Radio Astronomy Observatory.
The National Radio Astronomy Observatory is a facility of the National Science Foundation operated under cooperative agreement
by Associated Universities, Inc.}).
The \HI\ data cubes combine observations in the B, C, and D arrays and are characterized by
high sensitivity ($\le$1.1 mJy beam$^{-1}$ per channel), high spectral resolution (1.3 or 2.6 \kms), and
moderately high angular resolution (frequently, $\sim$6\arcsec).
In this study we used the robust-weighted
maps. In robust-weighting the spatial resolution is somewhat higher compared to naturally-weighted mapping and the beam shape is
more gaussian.
The galaxies used in this study and their properties that are useful here are listed in Table \ref{tab-gal}.

\begin{deluxetable}{lcccccccc}
\tabletypesize{\tiny}
\tablecaption{Sample Galaxies\label{tab-gal}}
\tablewidth{0pt}
\tablehead{
\colhead{} & \colhead{D} & \colhead{Beam size\tablenotemark{a}}
& \colhead{$\log \rm SFR_D^{FUV}$\tablenotemark{b}}
& \colhead{$\log M_{\rm HI~tot}$\tablenotemark{c}}
& \colhead{$R_{\rm Br}$\tablenotemark{d}}
& \colhead{$R_{\rm D}$\tablenotemark{d}}
& \colhead{$R_{\rm H\alpha}$\tablenotemark{d}}
& \colhead{$R_{\rm FUV}$\tablenotemark{d}} \\
\colhead{Galaxy} & \colhead{(Mpc)} & (pc)
& \colhead{($M$\solar yr$^{-1}$ kpc$^{-2}$)}
& \colhead{($M$\solar)} & \colhead{(kpc)}
& \colhead{(kpc)} & \colhead{(kpc)} & \colhead{(kpc)}
}
\startdata
CVnIdwA  &  3.6   &  187 & -2.48$\pm$0.01 & 7.67 &  0.56$\pm$0.49 &  0.25$\pm$0.12 & 0.69 &  0.49$\pm$0.03 \\
DDO 43      &  7.8 &  262 & -1.55$\pm$0.01 & 8.23 &  1.46$\pm$0.53 &  0.87$\pm$0.10 & 2.36 & 1.93$\pm$0.08 \\
DDO 46      &  6.1 &  170 & -2.46$\pm$0.01 & 8.27 &  1.27$\pm$0.18 &  1.13$\pm$0.05 & 1.51 & 3.02$\pm$0.06 \\
DDO 47      &  5.2 &  245 & -2.40$\pm$0.01 & 8.59 & \nodata              &  1.34$\pm$0.05 & 5.58  & 5.58$\pm$0.05 \\
DDO 50      &  3.4 &  108 & -1.55$\pm$0.01 & 8.85 &  2.65$\pm$0.27 &  1.48$\pm$0.06 & \nodata & 4.86$\pm$0.03 \\
DDO 52      & 10.3 &  297 & -2.43$\pm$0.01 & 8.43 &  2.80$\pm$1.35 &  1.26$\pm$0.04 & 3.69  & 3.39$\pm$0.10 \\
DDO 53      &  3.6  &  105 & -2.41$\pm$0.01 & 7.72 &  0.62$\pm$0.09 &  0.47$\pm$0.01 & 1.25 & 1.19$\pm$0.03 \\
DDO 63      &  3.9  &  130 & -1.95$\pm$0.00 & 8.19 &  1.31$\pm$0.10 &  0.68$\pm$0.01 & 2.26  & 2.89$\pm$0.04 \\
DDO 69      &  0.8  &    22 & -2.22$\pm$0.01 & 6.84 &  0.27$\pm$0.05 &  0.19$\pm$0.01 & 0.76  & 0.76$\pm$0.01 \\
DDO 70      &  1.3  &    85 & -2.16$\pm$0.00 & 7.61 &  0.13$\pm$0.07 &  0.44$\pm$0.01 & 1.23  & 1.34$\pm$0.01 \\
DDO 75      &  1.3  &    44 & -1.07$\pm$0.01 & 7.86 &  0.71$\pm$ 0.08 &  0.18$\pm$0.01 & 1.17 & 1.38$\pm$0.01 \\
DDO 87       &  7.7 &  256 & -1.00$\pm$0.01 & 8.39 &  0.99$\pm$0.11 &  1.21$\pm$0.02 & 3.18  & 4.23$\pm$0.07 \\
DDO 101     &  6.4 &  236 & -2.81$\pm$0.01 & 7.36 &  1.16$\pm$0.11 &  0.97$\pm$0.06 & 1.23  & 1.23$\pm$0.06 \\
DDO 126     &  4.9 &  147 & -2.10$\pm$0.01 & 8.16 &  0.60$\pm$0.05 &  0.84$\pm$0.13 & 2.84  & 3.37$\pm$0.05 \\
DDO 133     &  3.5 &  195 & -2.62$\pm$0.01 & 8.02 &  2.25$\pm$0.24 &  1.22$\pm$0.04 & 2.60 & 2.20$\pm$0.03 \\
DDO 154     &  3.7 &   127 & -1.93$\pm$0.01 & 8.46 &  0.62$\pm$0.09 &  0.48$\pm$0.02 & 1.73 & 2.65$\pm$0.04 \\
DDO 155     &  2.2 &   114 & \nodata               & 7.00 &  0.20$\pm$0.04 &  0.15$\pm$0.01 & 0.67 & \nodata \\
DDO 165     &  4.6 &   194 & \nodata              & 8.13 &  1.46$\pm$0.08 &  2.24$\pm$0.08 & 3.16 & \nodata \\
DDO 167     &  4.2 &   126 & -1.83$\pm$0.01 & 7.18 &  0.56$\pm$0.11 &  0.22$\pm$0.01 & 0.81 & 0.70$\pm$0.04 \\
DDO 168     &  4.3 &   141 & -2.04$\pm$0.01 & 8.45 &  0.72$\pm$0.07 &  0.83$\pm$0.01 & 2.24 & 2.25$\pm$0.04 \\
DDO 187     &  2.2 &     63 & -1.98$\pm$0.01 & 7.12 &  0.28$\pm$0.05 &  0.37$\pm$0.06 & 0.30 & 0.42$\pm$0.02 \\
DDO 210     &  0.9 &     44 & -2.71$\pm$0.06 & 6.30 & \nodata              &  0.16$\pm$0.01 & \nodata & 0.29$\pm$0.01 \\
DDO 216     &  1.1 &     84 & -3.21$\pm$0.01 & 6.75 &  1.77$\pm$0.45 &  0.52$\pm$0.01 & 0.42 & 0.59$\pm$0.01 \\
F564-V3      &  8.7 &    424 & -2.79$\pm$0.02 & 7.61 &  0.73$\pm$0.40 &  0.63$\pm$0.09 & \nodata & 1.24$\pm$0.08 \\
IC 10           &  0.7 &      19 & \nodata              & 7.78 &  0.30$\pm$0.04 &  0.39$\pm$0.01 & \nodata & \nodata \\
IC 1613       &  0.7 &      24 & -1.99$\pm$0.01 & 7.53 &  0.71$\pm$0.12 &  0.53$\pm$0.02 & \nodata & 1.77$\pm$0.01 \\
LGS 3          &  0.7 &     36 & -3.88$\pm$0.06 & 5.19 &  0.27$\pm$0.08 &  0.16$\pm$0.01 & \nodata & 0.32$\pm$0.01 \\
M81dwA       &  3.6 &   122 & -2.26$\pm$0.01 & 7.18 &  0.38$\pm$0.03 &  0.27$\pm$0.00 & \nodata & 0.71$\pm$0.03 \\
NGC 1569    &  3.4 &     91 & -0.01$\pm$0.01 & 8.39 &  0.85$\pm$0.24 &  0.46$\pm$0.02 & \nodata & 5.19$\pm$0.03 \\
NGC 2366    &  3.4 &    106 & -1.66$\pm$0.01 & 8.85 &  2.57$\pm$0.80 &  1.91$\pm$0.25 & 5.58 & 6.79$\pm$0.03 \\
NGC 3738    &  4.9 &    140 & -1.53$\pm$0.01 & 8.06 &  1.16$\pm$0.20 &  0.77$\pm$0.01 & 1.48 & 1.21$\pm$0.05 \\
NGC 4163    &  2.9 &    106 & -1.74$\pm$0.01 & 7.16 &  0.71$\pm$0.48 &  0.32$\pm$0.00 &  0.88 & 0.47$\pm$0.03 \\
NGC 4214    &  3.0 &    101 & -1.08$\pm$0.01 & 8.76 &  0.83$\pm$0.14 &  0.75$\pm$0.01 & \nodata & 5.46$\pm$0.03 \\
SagDIG        &  1.1 &    117 & -2.11$\pm$0.01 & 6.94 &  0.57$\pm$0.14 &  0.32$\pm$0.05 & 0.51 & 0.65$\pm$0.01 \\
UGC 8508    &  2.6 &      68 & \nodata             & 7.28 &  0.41$\pm$0.06 &  0.23$\pm$0.01 & 0.79 & \nodata \\
WLM             &  1.0 &      30 & -2.05$\pm$0.01 & 7.85 &  0.83$\pm$0.16 &  1.18$\pm$0.24 & 1.24 & 2.06$\pm$0.01 \\
Haro 29         &  5.8 &    173 & -1.07$\pm$0.01 & 7.80 &  1.15$\pm$0.26 &  0.33$\pm$0.00 & 0.96 & 0.86$\pm$0.06 \\
Haro 36         &  9.3 &    286 & -1.55$\pm$0.01 & 8.16 &  1.16$\pm$0.13 &  1.01$\pm$0.00 & 1.06 & 1.79$\pm$0.09 \\
Mrk178          &  3.9 &    110 & -1.66$\pm$0.01 & 6.99 &  0.38$\pm$0.00 &  0.19$\pm$0.00 & 1.17 & 1.45$\pm$0.04 \\
VII Zw 403     &  4.4 &    182 & -1.67$\pm$0.01 & 7.69 &  1.02$\pm$0.29 &  0.53$\pm$0.02 & 1.27 & 0.33$\pm$0.04 \\
\enddata
\tablenotetext{a}{Beam FWHM determined from the square-root of the major axis times the minor axis, for
robust-weighted maps \citep{lt}.}
\tablenotetext{b}{Integrated star formation rate (SFR) determined from the {\it Galaxy Evolution Explorer}
\citep[{\it GALEX};][]{galex}
far-ultraviolet (FUV) luminosity \citep{fuv}. The SFR is normalized by
dividing by the area within one disk scale length \rd.}
\tablenotetext{c}{Total \HI\ mass of the galaxy from \citet{lt}.}
\tablenotetext{d}{\rbr\ is the radius at which the $V$-band surface brightness profile changes slope, if it does \citep{breaks}.
\rd\ is the disk scale length determined from the $V$-band \citep{breaks}. \rha\ is the extent of \ha\ emission \citep{sfr}.
\rfuv\ is the radius of the furthest out FUV knot identified on {\it GALEX} images \citep{fuvknots}.
}
\end{deluxetable}

\section{Cloud Identification} \label{sec-id}

The cloud identifications were done with a program called CLOUDPROPS that runs under the Interactive Data Language (IDL) software \citep{cprops}.
CLOUDPROPS is a program originally built to identify molecular clouds by
decomposing emission into peaks and associated intensity clumps that appear to be unique in
position-velocity cubes. To use this program, we input the \HI\ data cube for the galaxy of
interest, the rms of individual \HI\ channel maps in the cube, and the distance to the galaxy.
We found that the program did not always locate clouds we could identify by
eye in the integrated \HI\ moment 0 map of the galaxy, so we experimented with four built-in, but optional, parameters in the code --
BMFRIENDS, SPECFRIEND, DELTA, and BCLIP -- to optimize them for our data.

\subsection{BMFRIENDS}

When searching for intensity peaks in the data cube, CLOUDPROPS identifies pixels with
intensities greater than all of the other surrounding pixels in a box.
BMFRIENDS or ``beam friends'' sets the length and width of this box.
The program's default for the length and width of the box is 3 beam sizes, BMFRIENDS=3.
Decreasing the size of the
box means there are more boxes in the space of the data cube, and if there are more boxes,
usually more clouds are identified.
This also affects the degree to which areas of emission are decomposed into smaller subregions.
When choosing a value for this parameter, we considered that
the same beam size covers a different physical area in a galaxy 1 Mpc away versus a galaxy 10 Mpc
away, assuming the absolute cloud sizes are similar between galaxies.

To test the effects of changing BMFRIENDS on the clouds, we compared the clouds
that the program found when we ran one galaxy through with a BMFRIENDS of 3, 4, 5, 6, 7 and 9. For
each value of BMFRIENDS, the program largely identified the same clouds but picked
up a few more clouds the lower the value of BMFRIENDS. For example,
in going from a BMFRIENDS of 5 to 4, we found that in two
cases the cloud peak was split into two smaller peaks. And in going from a BMFRIENDS of 5 to 6, we found that the
opposite occurred and two peaks were combined into one larger peak. However,
besides these two clouds, the clouds we found were reproducable for
a range of reasonable parameters and the changes primarily affected the identification of tinier clouds.

Therefore, we used BMFRIENDS=6 as the standard for a galaxy at a distance of about 3.6 Mpc, and
varied BMFRIENDS based on how far away other galaxies were:
a larger BMFRIENDS for closer galaxies and a smaller BMFRIENDS for more distant galaxies.

\subsection{SPECFRIENDS}

SPECFRIENDS sets the depth in velocity channels of the same box that BMFRIENDS sets the angular length and width for.
Because Hanning smoothing was applied during data processing,
we input the requirement that a cloud must appear in a minimum of 3 velocity channels (MINVCHAN=3),
but we also extended the SPECFRIENDS box.
The default for SPECFRIENDS is one velocity channel.
Extending the box had the consequence of decreasing the number of clouds found per galaxy,
but we found that SPECFRIENDS of 5 worked well and we used this value for SPECFRIENDS for all galaxies.

\subsection{DELTA}

DELTA designates how much a peak must
stand out above the noise and other gas in the galaxy to be selected.
As part of the iterative decomposition process, for
each pair of peaks, CLOUDPROPS defines the highest surface brightness level that includes both peaks as
the ``merge level''. As a default, the program will only consider
peaks that have an intensity that is 2$\times$rms above the merge level, which corresponds to
DELTA=2. Below 2$\times$rms, the decomposition is more likely to identify noise as cloud peaks. We
increased DELTA to 2.2, meaning the peak must be 2.2$\times$rms greater than the merge level, in
order to pick up less noise in the cloud identification step. We held DELTA constant for all galaxies.

\subsection{BCLIP}

BCLIP is used to vary how likely it is that
CLOUDPROPS will identify substructure within a cloud as individual,
smaller clouds.
BCLIP defaults to infinity such that no pixels are altered to reduce their contrast.
By setting BCLIP, the program decreases the contrast of pixels above a
specified intensity, assumes intensities above this level are associated with clouds, and uses the
reduced contrast data in the decomposition.
BCLIP is set in intensity units of the data, in our case Jy bm$^{-1}$.
Decreasing BCLIP has the effect of decreasing the number of clouds found.
We used BCLIP = 0.01 Jy bm$^{-1}$ because we found that this value generally neither clumped clouds
together that we felt should be separate nor separated substructure into individual clouds.

\subsection{Solving problems with CLOUDPROPS}

For both DDO 75 and DDO 133 the program entirely avoided particular large regions of the galaxy
where there were obvious clumps of gas.
We tried adjusting
the parameters and found none that would make the program touch those parts of the galaxy.
For these two galaxies, we cut out sub-cubes from the emission data cubes.
We ran these sub-cubes through CLOUDPROPS and this allowed the program to pick up the clouds in
each region. We then combined the clouds from the different searches into one catalog for each galaxy.

\subsection{Completeness}

\subsubsection{Comparison to finding clouds by hand}

The advantages to using an automatic algorithm for finding clouds are clear: no human biases and
much less time intensive.
With the parameters discussed above, CLOUDPROPS ran well, for the most part, on our \HI\ data cubes.
To see how identifying clouds by hand compares with the catalogues produced by CLOUDPROPS, we
identified clouds by hand in 10 galaxies from our sample. ``By hand'' means that we examined the \HI\ cube
channel by channel looking for coherent structures that were present over at least three channels.
The test galaxies were chosen to have at least
several clouds but not more than 20 to keep this test feasible.

\begin{figure}[t!]
\vskip -0.5truein
\includegraphics[scale=0.825]{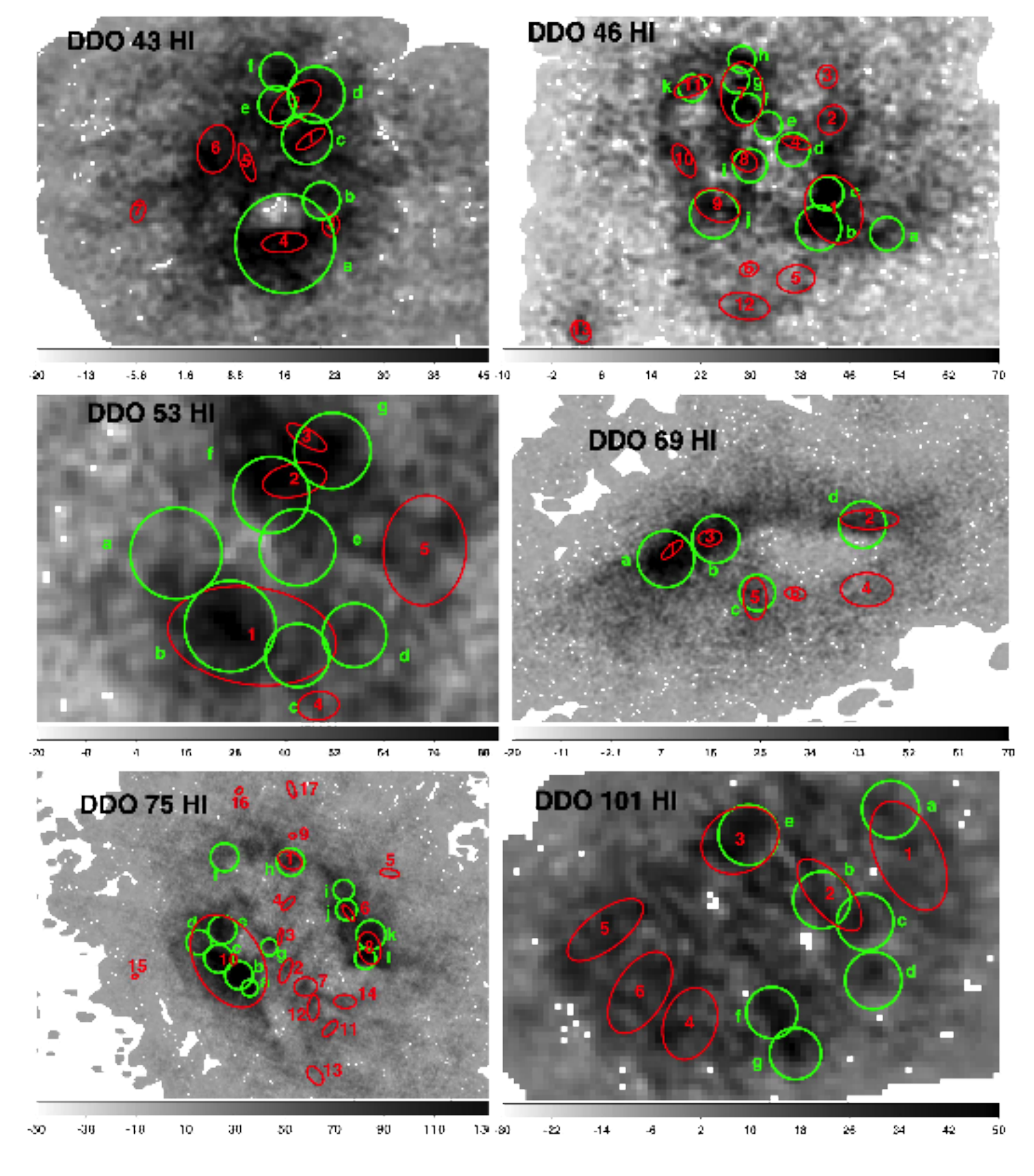}
\vskip -0.25truein
\caption{Clouds identified in 10 of our galaxies by hand (green) and by CLOUDPROPS (red) shown on
the integrated \HI\ moment 0 maps. Clouds identified by hand are outlined with a circle, whereas those
identified by CLOUDPROPS are ellipses defined by the moment 2 of the emission along the major and minor
axes of the cloud and the radius of the cloud is $R_{\rm cl}=1.91 \sqrt{\sigma_{maj} \sigma_{min}}$
(explained in detail in the next section).
Units of the gray scale bars below each image are Jy bm$^{-1}$ m s$^{-1}$.
\label{fig-byhandimage}}
\end{figure}

Figure \ref{fig-byhandimage} outlines the clouds we identified by hand and those identified by CLOUDPROPS
on the integrated \HI\ moment 0 maps of the galaxies.
We identified a total of 85 clouds by hand in the 10 galaxies, and CLOUDPROPS identified 74.
We missed 33 of the clouds that CLOUDPROPS identified and identified 26 clouds that CLOUDPROPS
had not identified. There were 13 occurrences of multiple clouds that we identified by hand being
parts of single CLOUDPROPS clouds, but only once were our clouds combined into a single cloud by
CLOUDPROPS.

\begin{figure}[t!]
\vskip -0.2truein
\includegraphics[scale=0.825]{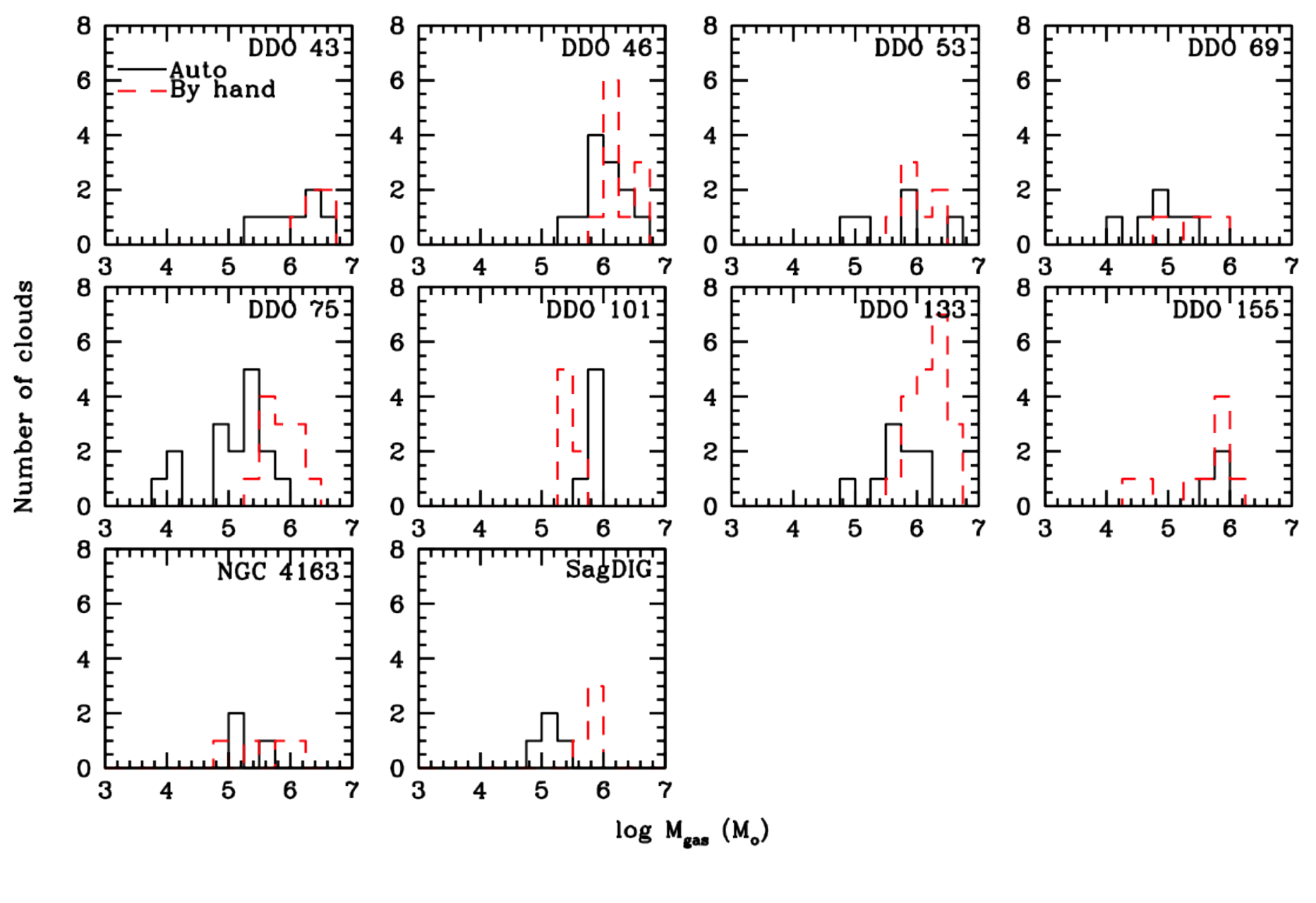}
\vskip -0.3truein
\caption{Histogram of cloud masses in our CLOUDPROPS catalogues for the 10 galaxies in our test of identifying clouds
by hand. The black histogram is the CLOUDPROPS catalog, and the red histogram is
the catalog of clouds identified by hand.
\label{fig-byhandhist}}
\end{figure}

Figure \ref{fig-byhandhist} shows a histogram of the cloud masses that were
identified by hand compared to the histogram of the clouds in the CLOUDPROPS
catalogues. (The masses of the clouds are discussed in more detail in the next
section).
The clouds identified by hand tend to be at the more massive end of the range of masses
found by CLOUDPROPS in several of the galaxies.

\clearpage
\includegraphics[scale=0.825]{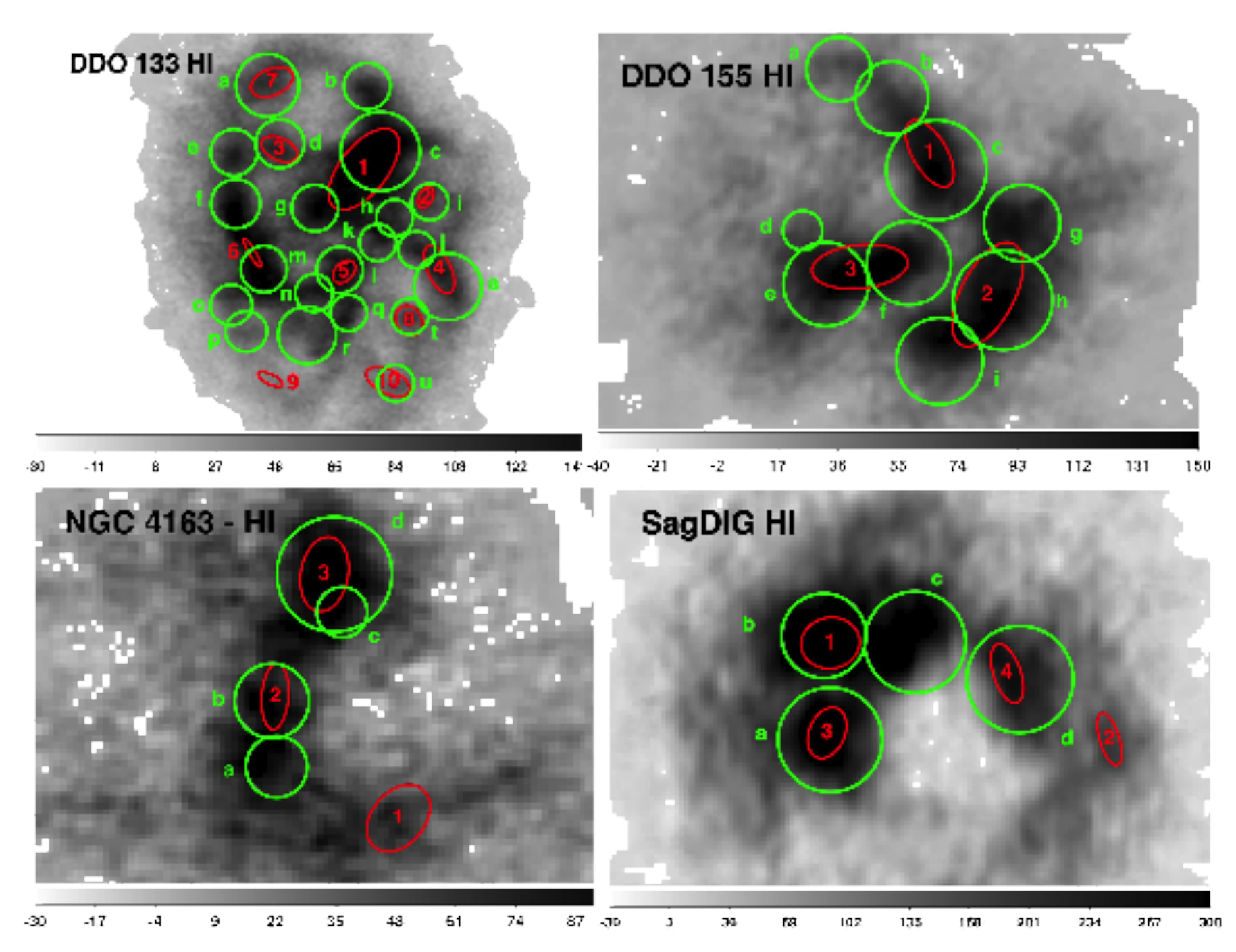}

\clearpage

\subsubsection{Finding added clouds}

To examine how well the program finds clouds for which we have prior knowledge, we added clouds found
in one galaxy to another galaxy and ran CLOUDPROPS with the original parameters for the host galaxy.
A cloud was added to a galaxy by summing the data cube of the galaxy being tested with the data cube of the
cloud. The data cube of the cloud was produced by using the CLOUDPROPS mask for that cloud to replace with
zero all pixels in the original galaxy data cube that do not belong to the cloud, producing a RA-Decl-Velocity
data cube of just the cloud.
Since different galaxies have different radial velocities and numbers of velocity channels,
as well as RA and Decl, we simply summed pixels with the velocity axis of the cloud centered on the central
velocity channel of the galaxy to which it was being added.

In particular we chose DDO 168 as the test host galaxy because it is of moderate size, at a distance (4.3 Mpc) not too different from
the median distance for our sample (3.6 Mpc), and has a
reasonable number of clouds of its own (21).
We added the clouds identified in DDO 154 (distance 3.7 Mpc and 34 clouds),
DDO 165 (distance 4.6 Mpc, 11 clouds), and NGC 2366 (distance 3.4 Mpc, 48 clouds)
in three separate runs of CLOUDPROPS.
We eliminated added clouds that fell outside the \HI\ disk of DDO 168 and
clouds that overlay DDO 168's own identified clouds.

\begin{figure}[t!]
\includegraphics[scale=0.825]{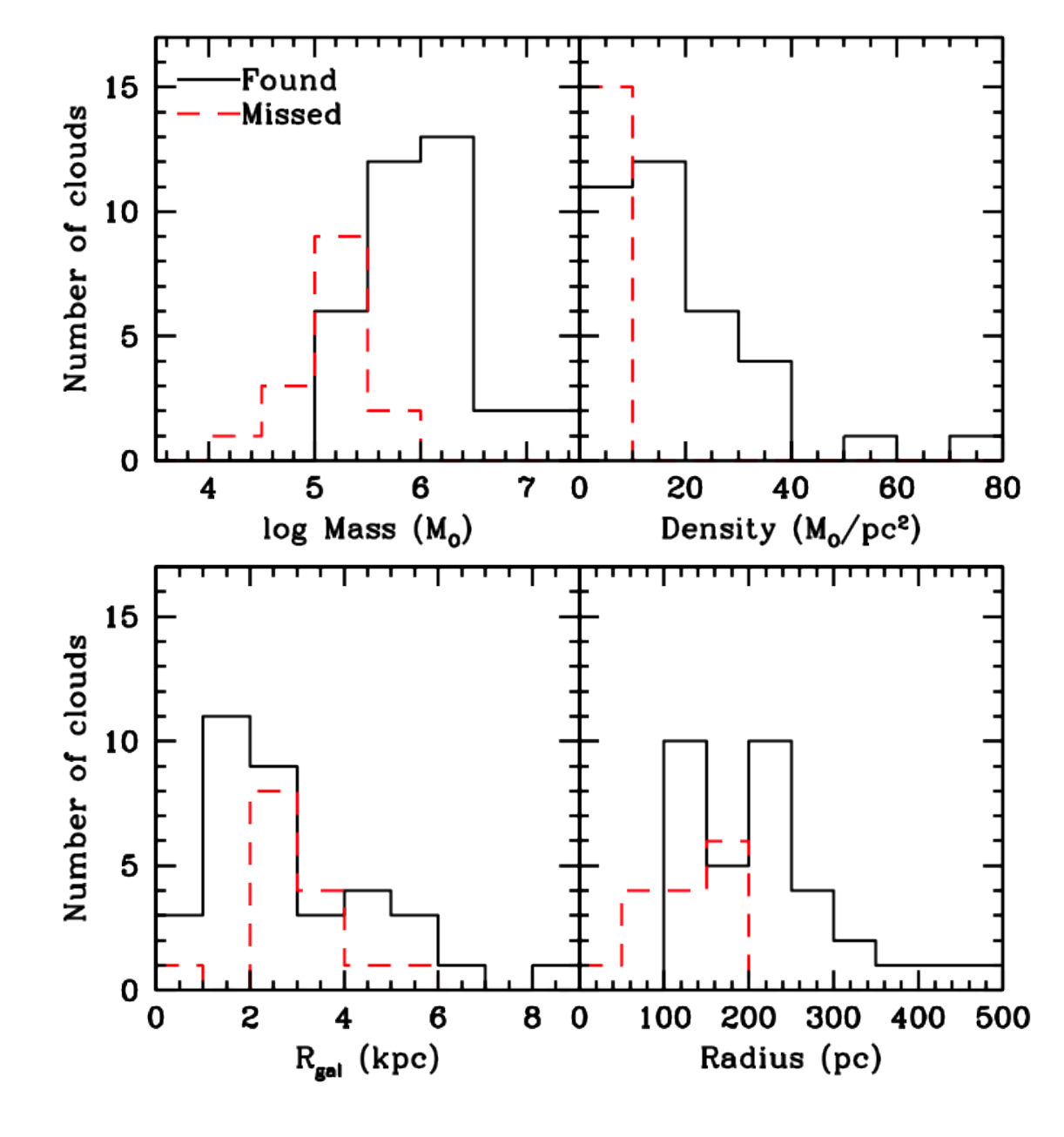}
\vskip -0.3truein
\caption{Histogram of mass, density, distance from the center of DDO 168, and radius
of missed and recovered clouds that were added to DDO 168. The added clouds were those identified
by CLOUDPROPS in DDO 154, DDO 165, and NGC 2366.
\label{fig-addedgals}}
\end{figure}

We ended with 15 added clouds that were missed and 35 added clouds that were found.
Histograms of the mass, density, distance from the center of DDO 168 in the plane of the galaxy, and radius
of the missed and found clouds are shown in Figure \ref{fig-addedgals}.
We see that missed clouds tend to be less massive than $10^6$ $M$\solar, have surface densities
lower than $10 M$\solar\ pc$^{-2}$, and be smaller than 200 pc in radius.
In terms of distance from the center of the galaxy, the missed clouds
are primarily found at middle radii.
The catalogues that are given here are unlikely to be complete except for the most massive clouds ($>10^6$ $M$\solar).
However, the program did pick out a representative sample of clouds from each galaxy.

\subsection{Extracting Properties from CLOUDPROPS}

For each galaxy, CLOUDPROPS outputs a catalog of clouds and their properties. The properties that we were interested
in included
position of the cloud in the galaxy; the central velocity $V_{\rm cen}$ and the velocity dispersion $V_{\rm disp}$;
the moment 2 of the emission along the major and minor axes of the cloud ($\sigma_{maj}$, $\sigma_{min}$);
the radius of the cloud; the position angle (P.\ A.) of the major axis of the cloud; the signal to noise of the peak intensity of the cloud;
and a mask that identifies the pixels in the \HI\ data cube that belong to the cloud.
All of our clouds have emission peaks with signal-to-noise greater than 4.

The rms spatial size determined by CLOUDPROPS from the moment 2 of the cloud emission is
$\sigma_r=\sqrt{\sigma_{maj} \sigma_{min}}$,
and the radius of the cloud is given by $R_{\rm cl}=1.91 \sigma_r$. The factor of 1.91 comes from \citet{solomon87}'s definition of
a cloud radius for giant molecular clouds.
Versions of the radius of the cloud were also given in which the radius is
deconvolved with the size of the beam, extrapolated to an edge temperature of 0 K, or both.
The extrapolation to an edge is more uncertain for clouds, like \HI\ clouds, that are sitting within extended emission than with molecular clouds
whose contrast with their surroundings may be higher.
We have chosen, therefore, to use radii without either deconvolution or extrapolation to 0 K, as did \citet{lmc} in their study of LMC clouds.
We use the CLOUDPROPS moment 2 size with its factor of 1.91
to convert to cloud radius, but note that \citet{lmc} used
$R_{\rm cl} = \sqrt{2 \ln2 \sigma_{maj} \sigma_{min}}$, where the factor $\sqrt{2 \ln2} = 1.18$
comes from one-half of the FWHM of a two-dimensional Gaussian distribution.

We examined a position-velocity (p-v) diagram for each cloud to make sure that it was separated in velocity space from other gas.
To make the p-v diagram, we summed the data cube over the Declination covered by the cloud.
We also used the mask generated by CLOUDPROPS as a
conditional transfer to produce a cube with non-cloud pixels blanked. We
summed the velocity channels for these cloud-specific cubes, to produce
a moment 0 image of each cloud.
We then integrated the total flux of each cloud and determined its mass $M_{\rm gas}$ and surface density $\Sigma_{\rm gas}$,
where the \HI\ mass is multiplied by 1.36 to include Helium. The surface density is the mass divided by the area of the ellipse
defined by \rcl\ and the minor-to-major axis ratio.
We also determined the distance of the cloud from the center of the galaxy in the plane of the galaxy $R_{\rm gal}$.
Finally, we determined the virial parameter
$\alpha_{\rm vir} = 5 \sigma_V^2 R / (G M) = 1163.1 (V_{\rm disp} ({\rm km/s}))^2 R_{\rm cl} ({\rm pc})/ M_{gas} (M\solar)$.

\begin{figure}[h!]
\vskip -1.15truein
\includegraphics[scale=0.7]{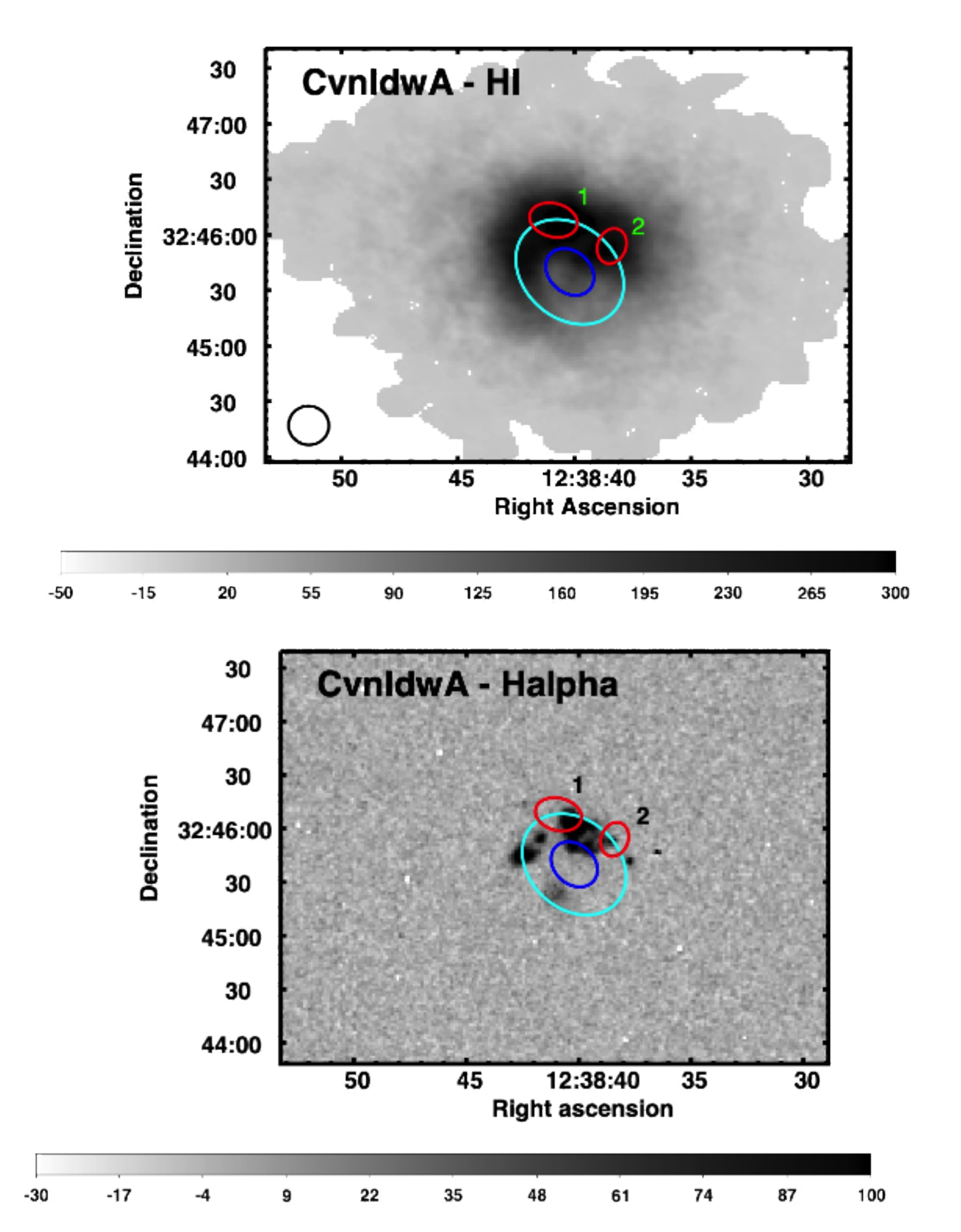}
\vskip -0.2truein
\caption{Clouds identified in CVnIdwA by CLOUDPROPS plotted as red ellipses on the integrated \HI\ map (top panel) and \ha\ image (bottom panel) of the galaxy.
The ellipse represents the size of the cloud as determined by CLOUDPROPS from the moment 2 of the emission along the major $\sigma_{maj}$ and minor $\sigma_{min}$
axes of the cloud, without deconvolution or extrapolation to an edge of 0 K.
The radius of the cloud is $R_{\rm cl}=1.91 \sqrt{\sigma_{maj} \sigma_{min}}$.
The small blue ellipse denotes a galactic radius of one disk scale length. The larger cyan ellipse outlines the radius at which the
$V$-band surface brightness profile of the galaxy changes slope, usually down-bending, if that occurs, taken from \citet{breaks}.
The black ellipse in the lower left corner of the \HI\ map outlines the FWHM of the beam,
and the units of the map are Jy bm$^{-1}$ m s$^{-1}$.
Figures for the rest of the galaxies in this study are available in the on-line materials.
\label{fig-cvnidwa}}
\end{figure}

\begin{figure}[h!]
\vskip -0.5truein
\includegraphics[scale=0.7]{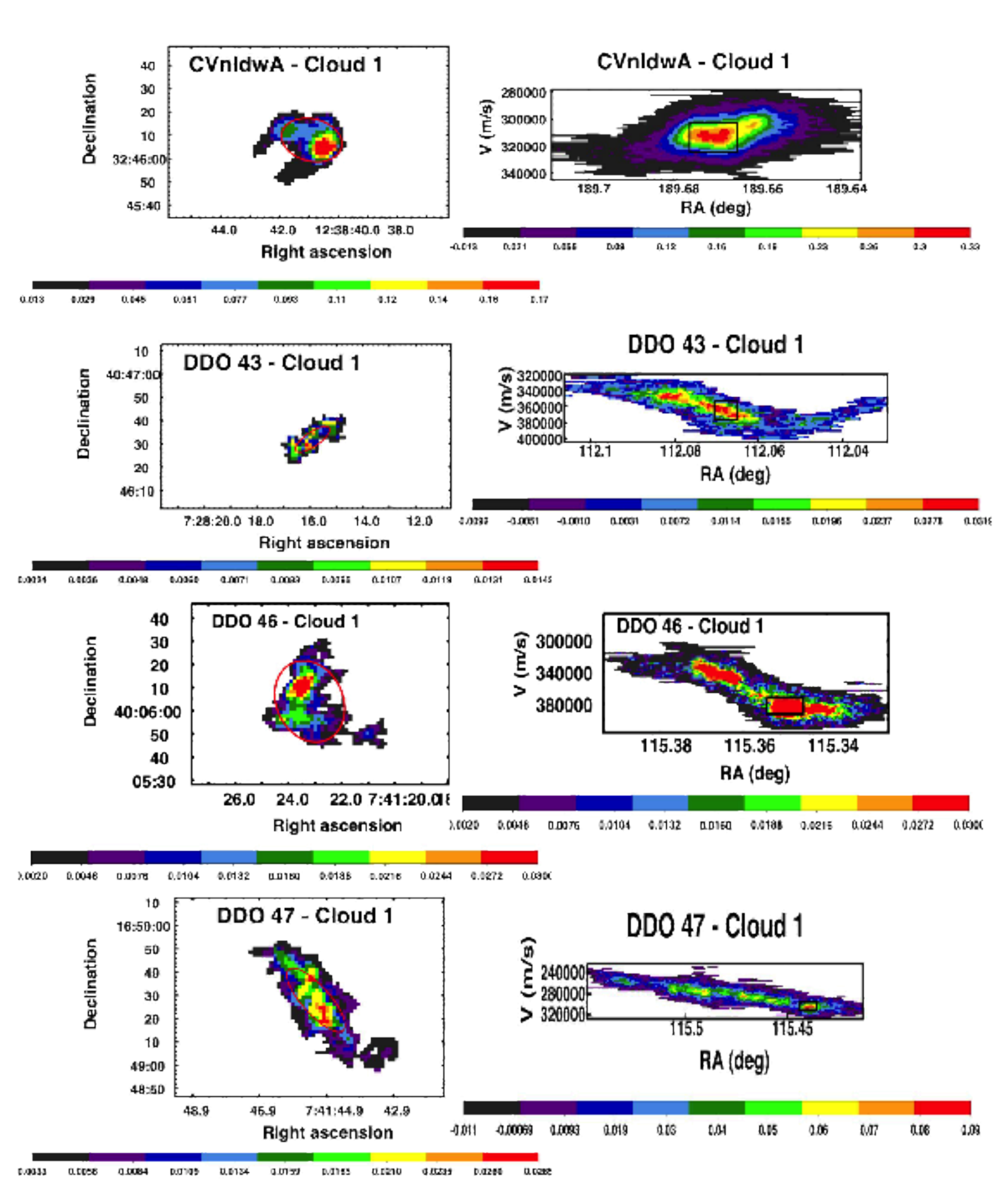}
\vskip -0.2truein
\caption{Moment 0 map and p-v diagram summed over Declination for cloud No.\ 1 in the catalog for each galaxy.
The red ellipse on the moment 0 map (left column) is the same as that in Figure \ref{fig-cvnidwa}, and the black or white box outlined
on the p-v diagram (right column) marks the extent of the cloud pixels in R. A. and velocity space
in the CLOUDPROPS cube mask that defines the cloud.
The units of the maps in the left column are Jy bm$^{-1}$ (summed over channels with units of m s$^{-1}$),
and the units of the channel maps in the right column are Jy bm$^{-1}$.
Figures for the rest of the galaxies in this study are available in the on-line materials.
\label{fig-clouds1}}
\end{figure}

The clouds identified in each galaxy are listed in Table \ref{tab-cat} and
outlined on the galaxy integrated \HI\ map and \ha\ image, if there is \ha\ emission,
in Figure \ref{fig-cvnidwa}. The ellipses denoting the clouds represent the major and minor axes and position angle as described above.
In Figure \ref{fig-clouds1} we include the cloud moment 0 and p-v diagram
of cloud No.\ 1 in each galaxy. Since the first cloud in the catalog is not first because of any property of the cloud itself,
this illustrates a range of cloud masses and sizes.
The red ellipse on each cloud in Figure \ref{fig-clouds1} is the same as that in Figure \ref{fig-cvnidwa}.

The uncertainty in the cloud masses given in Table \ref{tab-cat} are
calculated from the channel rms, ${\rm rms}_{\rm ch}$, given by \citet{lt} for
the \HI\ RA-Decl-Velocity data cubes. We multiply this by the square-root of the
number of summed velocity channels, $N_{ch}$, in the cloud and the square-root of the number
of beams in the cloud area, $N_{beam}$. 
The rms in mJy is then converted to solar mass by
multiplying by $235.6 \times {\rm D(Mpc)}^2 \times {\rm delV(km/s)}$, where delV
is the channel separation. 
Thus,
${\rm rms}_{M_{\rm gas}} = \sqrt{(N_{ch} \times N_{beam})} \times {\rm rms}_{\rm ch} \times 235.6 \times {\rm D^2} \times {\rm delV}$
and
${\rm rms}_{\log M_{\rm gas}} = 1/2.3026 \times {\rm rms}_{M_{\rm gas}}/M_{\rm gas}$.

The uncertainty in the gas density, ${\rm rms}_{\Sigma_{\rm gas}}$, is derived from the 
uncertainty in the mass and the uncertainty in $R_{cl}$:
$\Sigma_{\rm gas} = M_{\rm gas}/(\pi \times R_{\rm cl}^2 \times (\sigma_{\rm min}/\sigma_{\rm maj}))$
and
$({\rm rms}_{\Sigma_{\rm gas}}/\Sigma_{\rm gas})^2 = ({\rm rms}_{M_{\rm gas}}/M_{\rm gas})^2 + ({\rm rms}_{R_{cl}}/R_{cl})^2
+ ({\rm rms}_{\rm rat}/{\rm rat})^2$, where ${\rm rat} = \sigma_{\rm min}/\sigma_{\rm maj}$.
The uncertainty in $\alpha_{vir}$, ${\rm rms}_{\alpha_{vir}}$, is derived from the uncertainties in the velocity dispersion,
$R_{cl}$, and mass:
$\alpha_{vir} = 1163.1 \times V_{\rm disp}^2 \times R_{\rm cl} /  M_{gas}$
and
$({\rm rms}_{\alpha_{vir}}/\alpha_{vir})^2 = (2 \times {\rm rms}_{V_{\rm disp}}/V_{\rm disp})^2 + 
({\rm rms}_{R_{cl}}/R_{cl})^2+ ({\rm rms}_{M_{\rm gas}}/M_{\rm gas})^2$.

\clearpage

\begin{deluxetable}{lrcccrcccccccc}
\tabletypesize{\tiny}
\rotate
\tablecaption{Cloud Catalog\label{tab-cat}}
\tablewidth{0pt}
\tablehead{
\colhead{} & \colhead{}
& \colhead{R.\ A.}
& \colhead{Decl.}
& \colhead{$V_{\rm cen}$}
& \colhead{P.\ A.}
& \colhead{$R_{\rm cl}$}
& \colhead{}
& \colhead{}
& \colhead{$V_{\rm disp}$}
& \colhead{$\log M_{\rm gas}$}
& \colhead{$\Sigma_{\rm gas}$}
& \colhead{$R_{\rm gal}$}
& \colhead{} \\
\colhead{Galaxy} & \colhead{Cloud}
& \colhead{(h m s)}
& \colhead{(d m s)}
& \colhead{(\kms)}
& \colhead{(deg)}
& \colhead{(pc)}
& \colhead{$\sigma_{\rm min}/\sigma_{\rm maj}$}
& \colhead{Peak S/N}
& \colhead{(\kms)}
& \colhead{(M\solar)}
& \colhead{(M\solar pc$^{-2}$)}
& \colhead{(kpc)}
& \colhead{$\alpha_{\rm vir}$}
}
\startdata
CVnIdwA  &   1 & 12 38 40.9 &  32 46 08 &  312.4 &   76.2 &  229$\pm$21 & 0.69 & 24.3 &   2.9$\pm$0.3 & 6.25$\pm$0.01 &  15.6$\pm$2.1 &  0.6 &   1.3$\pm$0.1 \\
         &   2 & 12 38 38.4 &  32 45 54 &  301.7 &  -21.2 &  170$\pm$19 & 0.78 & 25.5 &   2.9$\pm$0.3 & 6.02$\pm$0.01 &  14.7$\pm$2.5 &  0.6 &   1.6$\pm$0.2 \\
DDO 43   &   1 &  7 28 15.9 &  40 46 32 &  366.6 &  -55.6 &  329$\pm$36 & 0.35 &  7.4 &   4.9$\pm$1.0 & 6.21$\pm$0.02 &  13.4$\pm$2.3 &  1.2 &   5.7$\pm$0.7 \\
         &   2 &  7 28 16.6 &  40 46 50 &  364.0 &  -49.9 &  591$\pm$53 & 0.52 &  7.4 &   6.5$\pm$0.7 & 6.39$\pm$0.03 &   4.3$\pm$0.7 &  1.9 &  11.8$\pm$1.4 \\
         &   3 &  7 28 15.0 &  40 45 48 &  361.7 &   -4.1 &  219$\pm$53 & 0.73 &  7.2 &   1.1$\pm$0.8 & 5.47$\pm$0.05 &   2.6$\pm$1.0 &  1.7 &   1.1$\pm$0.3 \\
         &   4 &  7 28 17.1 &  40 45 39 &  355.6 &  -84.3 &  436$\pm$39 & 0.44 &  8.5 &   3.9$\pm$0.4 & 6.64$\pm$0.01 &  16.5$\pm$2.3 &  1.5 &   1.8$\pm$0.2 \\
         &   5 &  7 28 18.8 &  40 46 20 &  352.2 &   18.9 &  384$\pm$62 & 0.31 &  6.2 &   2.4$\pm$0.5 & 5.91$\pm$0.03 &   5.6$\pm$1.4 &  0.8 &   3.2$\pm$0.6 \\
         &   6 &  7 28 20.2 &  40 46 27 &  346.2 &  -15.5 &  473$\pm$52 & 0.73 &  6.5 &   3.6$\pm$0.7 & 6.35$\pm$0.03 &   4.4$\pm$0.8 &  1.6 &   3.2$\pm$0.4 \\
         &   7 &  7 28 23.7 &  40 45 55 &  337.7 &  -15.5 &  214$\pm$36 & 0.61 &  5.4 &   2.1$\pm$0.6 & 5.61$\pm$0.04 &   4.6$\pm$1.3 &  2.7 &   2.7$\pm$0.5 \\
DDO 46   &   1 &  7 41 23.2 &  40 06 04 &  385.6 &   24.6 &  537$\pm$32 & 0.79 &  9.7 &   5.3$\pm$0.5 & 6.95$\pm$0.01 &  12.5$\pm$1.2 &  1.5 &   2.0$\pm$0.1 \\
         &   2 &  7 41 23.3 &  40 06 50 &  390.9 &  -44.2 &  233$\pm$28 & 0.85 &  5.6 &   5.6$\pm$1.1 & 5.96$\pm$0.04 &   6.2$\pm$1.3 &  1.0 &   9.4$\pm$1.4 \\
         &   3 &  7 41 23.5 &  40 07 12 &  380.6 &    1.7 &  172$\pm$26 & 0.89 &  4.8 &   3.5$\pm$0.7 & 5.70$\pm$0.04 &   6.0$\pm$1.5 &  1.4 &   5.0$\pm$0.9 \\
         &   4 &  7 41 24.9 &  40 06 38 &  377.0 &   76.2 &  223$\pm$49 & 0.35 &  6.4 &   3.6$\pm$0.7 & 5.92$\pm$0.02 &  15.2$\pm$5.1 &  0.4 &   4.0$\pm$0.9 \\
         &   5 &  7 41 24.9 &  40 05 28 &  368.0 &  -84.3 &  290$\pm$29 & 0.71 &  5.1 &   3.5$\pm$0.7 & 5.87$\pm$0.04 &   3.9$\pm$0.7 &  2.4 &   5.6$\pm$0.8 \\
         &   6 &  7 41 27.0 &  40 05 33 &  356.1 &  -72.8 &  143$\pm$31 & 0.73 &  5.8 &   3.1$\pm$0.6 & 5.35$\pm$0.07 &   4.8$\pm$1.8 &  2.2 &   7.1$\pm$1.9 \\
         &   7 &  7 41 27.3 &  40 07 03 &  353.4 &    1.7 &  482$\pm$24 & 0.68 &  8.1 &   9.0$\pm$0.9 & 6.74$\pm$0.01 &  11.2$\pm$0.9 &  0.9 &   8.2$\pm$0.5 \\
         &   8 &  7 41 27.2 &  40 06 29 &  348.4 &   59.0 &  210$\pm$21 & 0.69 &  7.2 &   5.4$\pm$0.5 & 6.15$\pm$0.02 &  14.8$\pm$2.4 &  0.5 &   5.0$\pm$0.6 \\
         &   9 &  7 41 28.4 &  40 06 06 &  343.2 &   70.4 &  353$\pm$39 & 0.67 &  6.6 &   4.5$\pm$0.5 & 6.31$\pm$0.02 &   7.7$\pm$1.3 &  1.3 &   4.1$\pm$0.5 \\
         &  10 &  7 41 29.9 &  40 06 29 &  329.1 &   30.3 &  273$\pm$33 & 0.50 &  6.1 &   4.2$\pm$0.4 & 6.11$\pm$0.02 &  11.0$\pm$2.1 &  1.3 &   4.3$\pm$0.6 \\
         &  11 &  7 41 29.5 &  40 07 07 &  335.7 &  -67.1 &  306$\pm$31 & 0.43 &  7.5 &   3.8$\pm$0.8 & 6.20$\pm$0.02 &  12.4$\pm$1.9 &  1.5 &   3.2$\pm$0.3 \\
         &  12 &  7 41 27.2 &  40 05 14 &  357.7 &   81.9 &  380$\pm$30 & 0.53 &  5.5 &   4.9$\pm$0.5 & 6.39$\pm$0.02 &  10.1$\pm$1.3 &  2.9 &   4.3$\pm$0.4 \\
         &  13 &  7 41 34.5 &  40 05 01 &  314.8 &   30.3 &  174$\pm$21 & 0.78 &  5.2 &   5.5$\pm$1.1 & 5.91$\pm$0.03 &  10.9$\pm$2.1 &  4.3 &   7.6$\pm$1.1 \\
\enddata
\tablecomments{Table \ref{tab-cat} is published in its entirety in the electronic edition of the journal. A portion is shown
here for guidance regarding its form and content.}
\end{deluxetable}

\section{Results} \label{sec-results}

\subsection{Cloud Properties}

\begin{figure}[t!]
\epsscale{1.0}
\vskip -1.0truein
\plotone{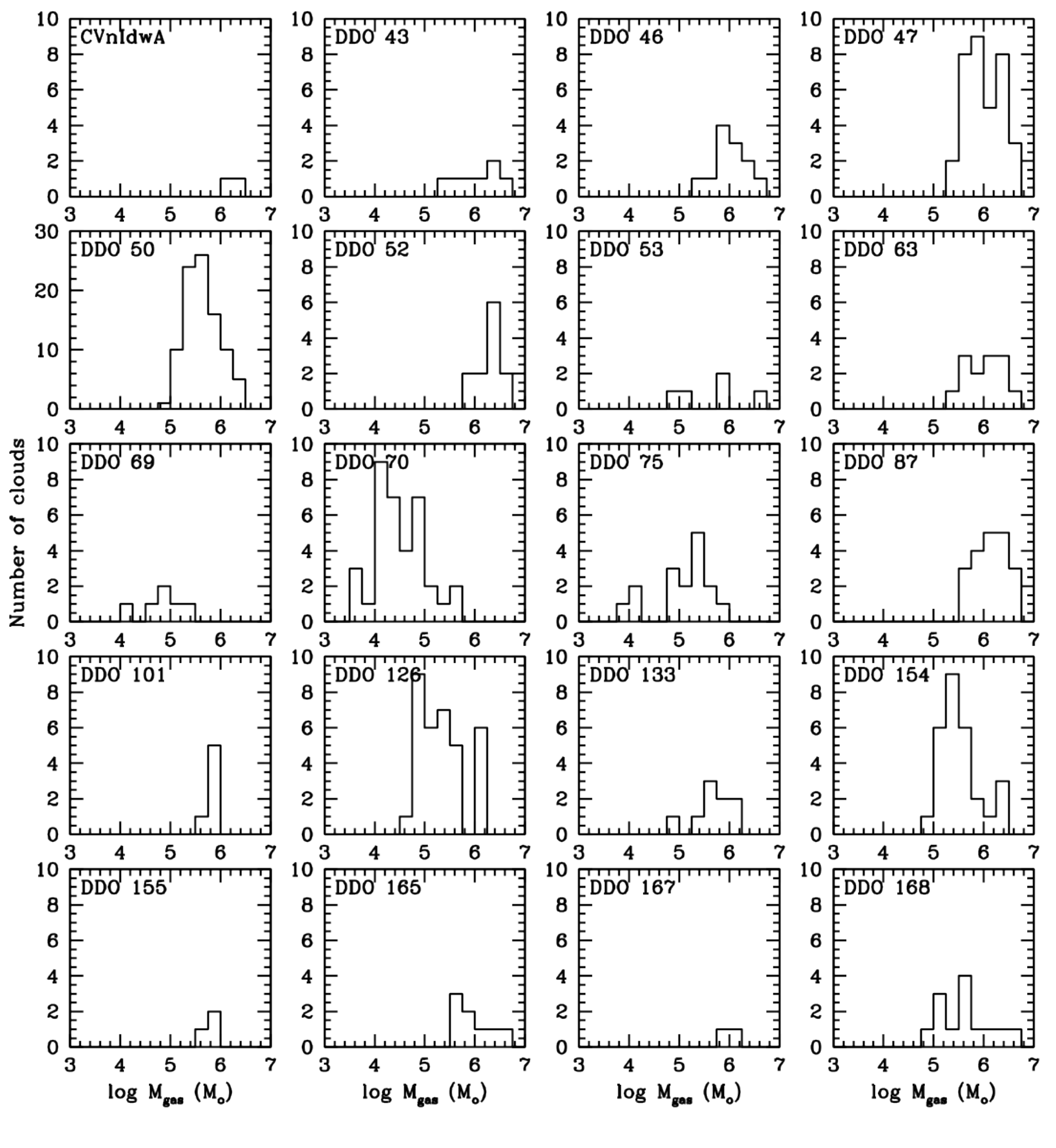}
\vskip -0.25truein
\caption{Number of clouds as a function of cloud gas mass for each galaxy.
\label{fig-histmass}}
\end{figure}

Table \ref{tab-cat} gives the properties of the individual clouds including position, central velocity, position angle,
major axis radius $R_{\rm cl}$, minor-to-major axis ratio, signal-to-noise (S/N) of the peak, velocity
dispersion $V_{\rm disp}$, mass of gas (\HI\ plus He), gas surface density $\Sigma_{gas}$,
radius from the center of the galaxy in the plane of the galaxy $R_{\rm gal}$, and the virial ratio $\alpha_{\rm vir}$.
There are 814
clouds in 40 galaxies, and individual galaxies have from one cloud (DDO 210, M81dwA) to 161 clouds (NGC 4214).
The number of clouds identified in each galaxy are given in Table \ref{tab-global}.
Cloud mass functions are plotted as histograms in Figure \ref{fig-histmass}.

\begin{deluxetable}{lcccccccc}
\tabletypesize{\tiny}
\tablecaption{Galaxy Cloud Properties\label{tab-global}}
\tablewidth{0pt}
\tablehead{
\colhead{}
& \colhead{} & \colhead{$\log M_{\rm clouds}$}
& \colhead{} & \colhead{$\log M_{\rm 3rd~cloud}$\tablenotemark{a}}
& \colhead{Fraction\tablenotemark{b}} & \colhead{Fraction\tablenotemark{b}}
& \colhead{Fraction\tablenotemark{b}} & \colhead{Fraction\tablenotemark{b}} \\
\colhead{Galaxy} & \colhead{No.\ Clouds}
& \colhead{($M$\solar)}
& \colhead{$M_{\rm clouds}/M_{\rm HI~tot}$}
& \colhead{($M$\solar)}
& \colhead{beyond $R_{\rm Br}$}
& \colhead{beyond $R_{\rm D}$}
& \colhead{beyond $R_{\rm H\alpha}$}
& \colhead{beyond $R_{\rm FUV}$}
}
\startdata
CVnIdwA  &   2 & 6.32 & 0.05 & \nodata &  1.00 & 1.00 &  0.00 &  1.00 \\
DDO 43      &   7 & 6.95 & 0.05 &   6.35 &  0.71 & 0.86 &  0.14 &  0.14 \\
DDO 46      &  13 & 7.30 & 0.11 &   6.39 &  0.69 & 0.69 &  0.38 &  0.08 \\
DDO 47      &  35 & 7.57 & 0.09 &   6.50 & \nodata & 0.94 &  0.17 &  0.17 \\
DDO 50      &  92 & 7.60 & 0.06 &   6.33 &  0.72 & 0.87 & \nodata &  0.21 \\
DDO 52      &  13 & 7.42 & 0.10 &   6.59 &  0.38 & 0.85 &  0.15 &  0.31 \\
DDO 53      &   5 & 6.70 & 0.10 &   5.81 &  0.20 & 0.80 &  0.00 &  0.00 \\
DDO 63      &  13 & 7.13 & 0.09 &   6.28 &  0.38 & 1.00 &  0.00 &  0.00 \\
DDO 69      &   6 & 5.66 & 0.06 &   4.91 &  1.00 & 1.00 &  0.17 &  0.17 \\
DDO 70      &  36 & 6.27 & 0.05 &   5.38 &  1.00 & 0.94 &  0.64 &  0.58 \\
DDO 75      &  17 & 7.17 & 0.20 &   5.68 &  0.71 & 0.94 &  0.35 &  0.24 \\
DDO 87      &  20 & 7.42 & 0.11 &   6.59 &  1.00 & 0.90 &  0.55 &  0.35 \\
DDO 101     &   6 & 6.53 & 0.15 &   5.94 &  0.67 & 0.83 &  0.50 &  0.50 \\
DDO 126     &  34 & 6.97 & 0.07 &   6.08 &  0.94 & 0.94 &  0.26 &  0.15 \\
DDO 133     &  10 & 7.12 & 0.12 &   6.05 &  0.30 & 0.80 &  0.10 &  0.30 \\
DDO 154     &  28 & 7.09 & 0.04 &   6.40 &  0.93 & 0.96 &  0.71 &  0.50 \\
DDO 155     &   3 & 6.25 & 0.18 &   5.99 &  1.00 & 1.00 &  0.00 & \nodata \\
DDO 165     &   9 & 7.18 & 0.11 &   6.35 &  0.78 & 0.33 &  0.11 & \nodata \\
DDO 167     &   2 & 6.25 & 0.12 & \nodata &  0.50 & 1.00 &  0.00 &  0.00 \\
DDO 168     &  14 & 7.27 & 0.07 &   6.41 &  0.93 & 0.93 &  0.43 &  0.43 \\
DDO 187     &   2 & 5.82 & 0.05 & \nodata &  0.50 & 0.00 &  0.50 &  0.00 \\
DDO 210     &   1 & 6.02 & 0.53 & \nodata & \nodata & 0.00 & \nodata &  0.00 \\
DDO 216     &   3 & 5.80 & 0.11 &   5.40 &  0.00 & 0.33 &  0.67 &  0.00 \\
F564-V3   &   4 & 6.11 & 0.03 &   5.65 &  1.00 & 1.00 & \nodata &  0.50 \\
IC 10     &  69 & 6.33 & 0.04 &   5.33 &  0.96 & 0.88 & \nodata & \nodata \\
IC 1613   &  93 & 6.46 & 0.09 &   5.33 &  0.94 & 0.98 & \nodata &  0.41 \\
LGS 3     &   4 & 3.89 & 0.05 &   3.24 &  0.00 & 0.25 & \nodata &  0.00 \\
M81dwA   &   1 & 5.49 & 0.02 & \nodata &  1.00 & 1.00 & \nodata &  0.00 \\
NGC 1569    &  26 & 6.78 & 0.02 &   5.87 &  0.81 & 0.92 & \nodata &  0.00 \\
NGC 2366    &  44 & 7.58 & 0.05 &   6.53 &  0.80 & 0.84 &  0.09 &  0.00 \\
NGC 3738    &   6 & 7.17 & 0.13 &   5.70 &  1.00 & 1.00 &  1.00 &  1.00 \\
NGC 4163    &   3 & 5.64 & 0.03 &   5.06 &  0.00 & 0.67 &  0.00 &  0.33 \\
NGC 4214    & 161 & 7.73 & 0.09 &   6.52 &  0.98 & 0.99 & \nodata &  0.17 \\
SagDIG   &   4 & 5.65 & 0.05 &   5.22 &  0.50 & 1.00 &  0.75 &  0.25 \\
UGC 8508    &   3 & 6.27 & 0.10 &   5.96 &  0.67 & 0.67 &  0.33 & \nodata \\
WLM      &  11 & 6.23 & 0.02 &   5.44 &  0.82 & 0.64 &  0.46 &  0.09 \\
Haro 29   &   3 & 6.18 & 0.02 &   5.59 &  0.67 & 0.67 &  0.67 &  0.67 \\
Haro 36   &   5 & 7.23 & 0.12 &   6.53 &  0.40 & 0.40 &  0.40 &  0.20 \\
Mrk 178   &   2 & 6.26 & 0.18 & \nodata &  1.00 & 1.00 &  0.00 &  0.00 \\
VII Zw 403 &   4 & 6.59 & 0.08 &   5.22 &  0.50 & 0.75 &  0.50 &  1.00 \\
\enddata
\tablenotetext{a}{\HI\ mass of third most massive cloud for galaxies with at least three clouds.}
\tablenotetext{b}{Fraction of clouds beyond the extent of \rbr, \rd, \rha, and \rfuv.}
\end{deluxetable}

\clearpage

~~~~~~~
~~~~~~~

\includegraphics[scale=0.85]{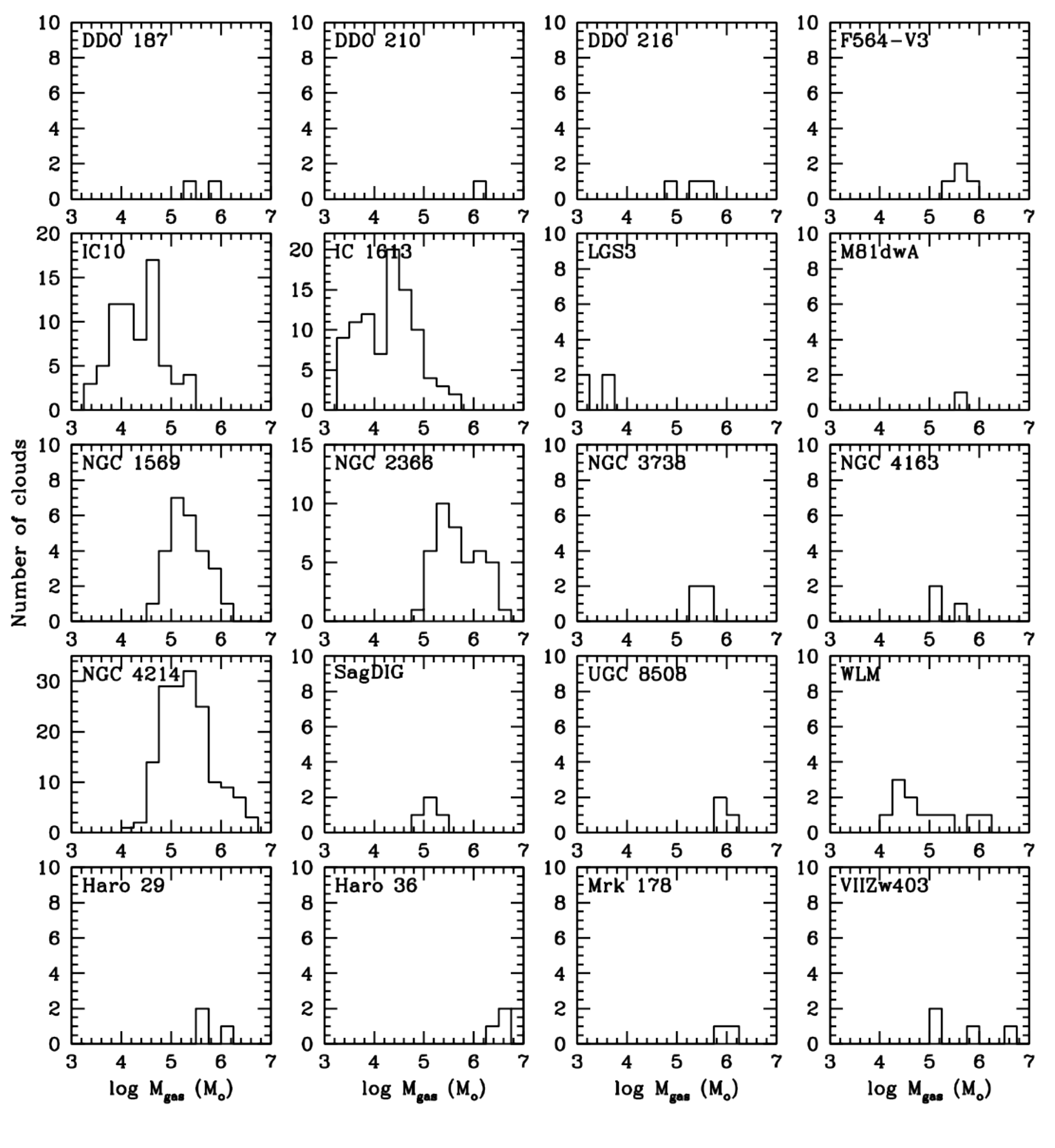}

\clearpage

\begin{figure}[t!]
\epsscale{1.2}
\plottwo{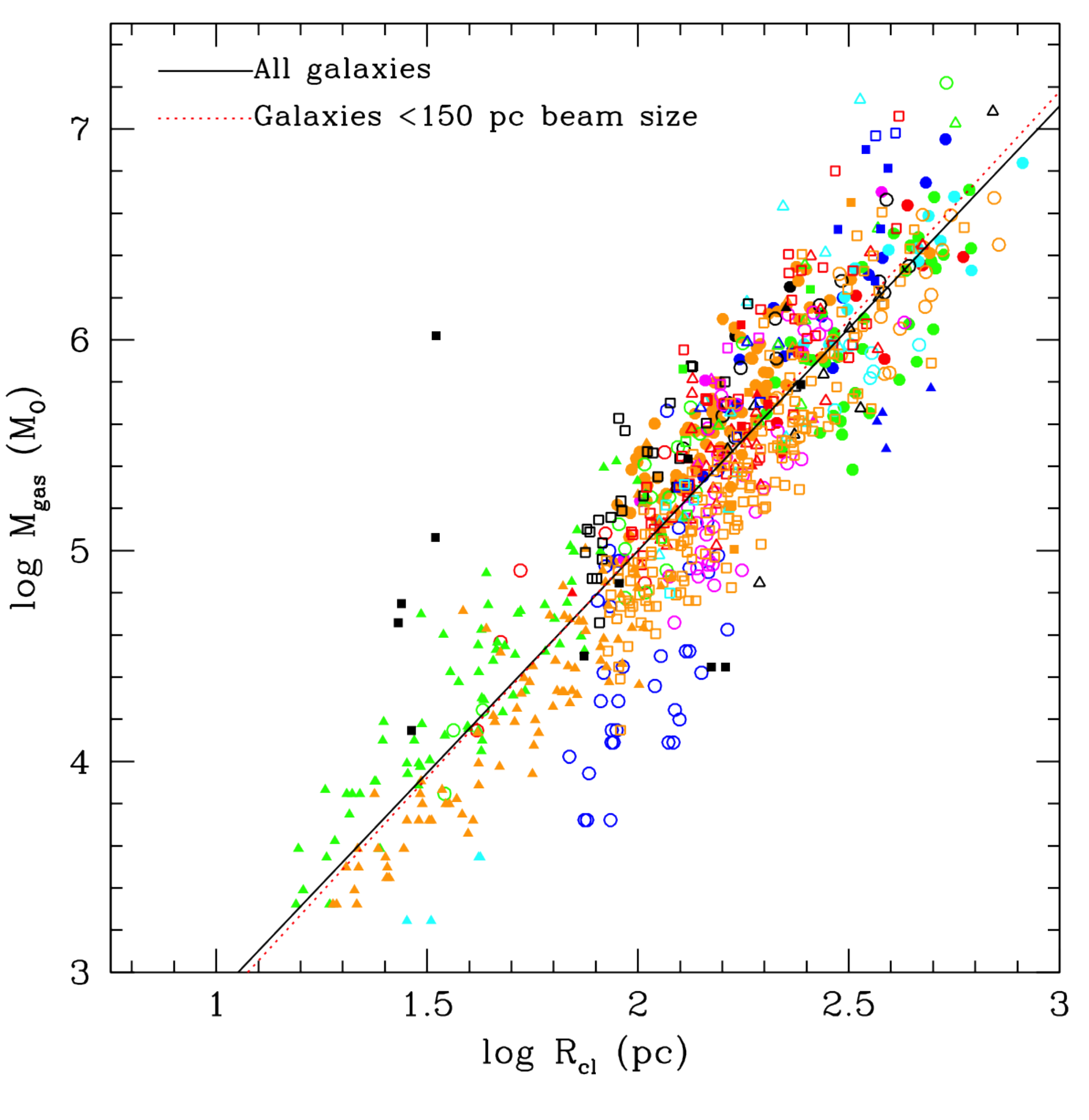}{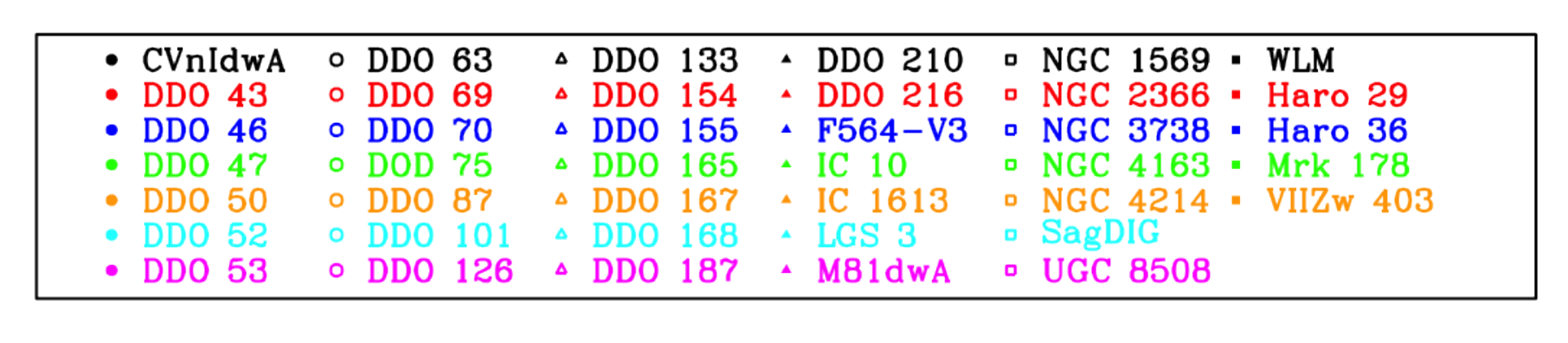}
\caption{Cloud gas mass as a function of cloud radius.
The solid black line is a fit to all of our clouds; the dotted red line is a fit to clouds only
in galaxies with beam sizes $\le150$ pc.
\label{fig-massvssize}}
\end{figure}

In Figure \ref{fig-massvssize} we plot the mass of the cloud against the size of the cloud.
\citet{lmc} surveyed the LMC and identified 468 \HI\ clouds in that galaxy.
The cloud masses identified by \citet{lmc} vary from $10^3$ $M$\solar\ to $10^5$ M\solar\
and ours vary from $10^3$ $M$\solar\ to $10^7$ M\solar.
The LMC radii are mostly between 10 pc and 200 pc for a brightness threshold of 16 K (their Figure 11)
while ours vary from 15 pc to 800 pc.
The biggest cloud in our study is found in DDO 52 (820 pc), the most distant galaxy in our sample with a beam size
of 300 pc, and the second and third largest clouds (720 pc, 700 pc) are found in DDO 87 with a beam size of 256 pc.
This is what one would expect if clouds are merging into larger complexes at the lower spatial resolution of more distant galaxies.
However, the fourth largest cloud (695 pc) is found in DDO 133 at 3.5 Mpc with a beam size of 195 pc,
and a cloud of radius 540 pc is found in DDO 75 with a beam size of only 44 pc.

We see that generally the larger the cloud, the more massive it is,
although we see that the clouds in DDO 70 (open blue
circles) do fall systematically to lower gas mass for a given radius compared to the rest of the galaxies.
The solid
black line is a fit to all 814 of our clouds: $\log M_{\rm gas} =
(2.11\pm0.04)\times \log R_{\rm cl} + (0.78\pm0.08)$. The dotted red line is a
fit to the 643 clouds in galaxies with beam sizes $\le 150$ pc: $\log M_{\rm gas}
= (2.17\pm0.04)\times \log R_{\rm cl} + (0.67\pm0.09)$.
The relationship found by \citet{lmc} for clouds in the LMC
has a slope of 2.05, which is similar to our slope of 2.11$\pm$0.04.

\begin{figure}[t!]
\epsscale{0.5}
\plotone{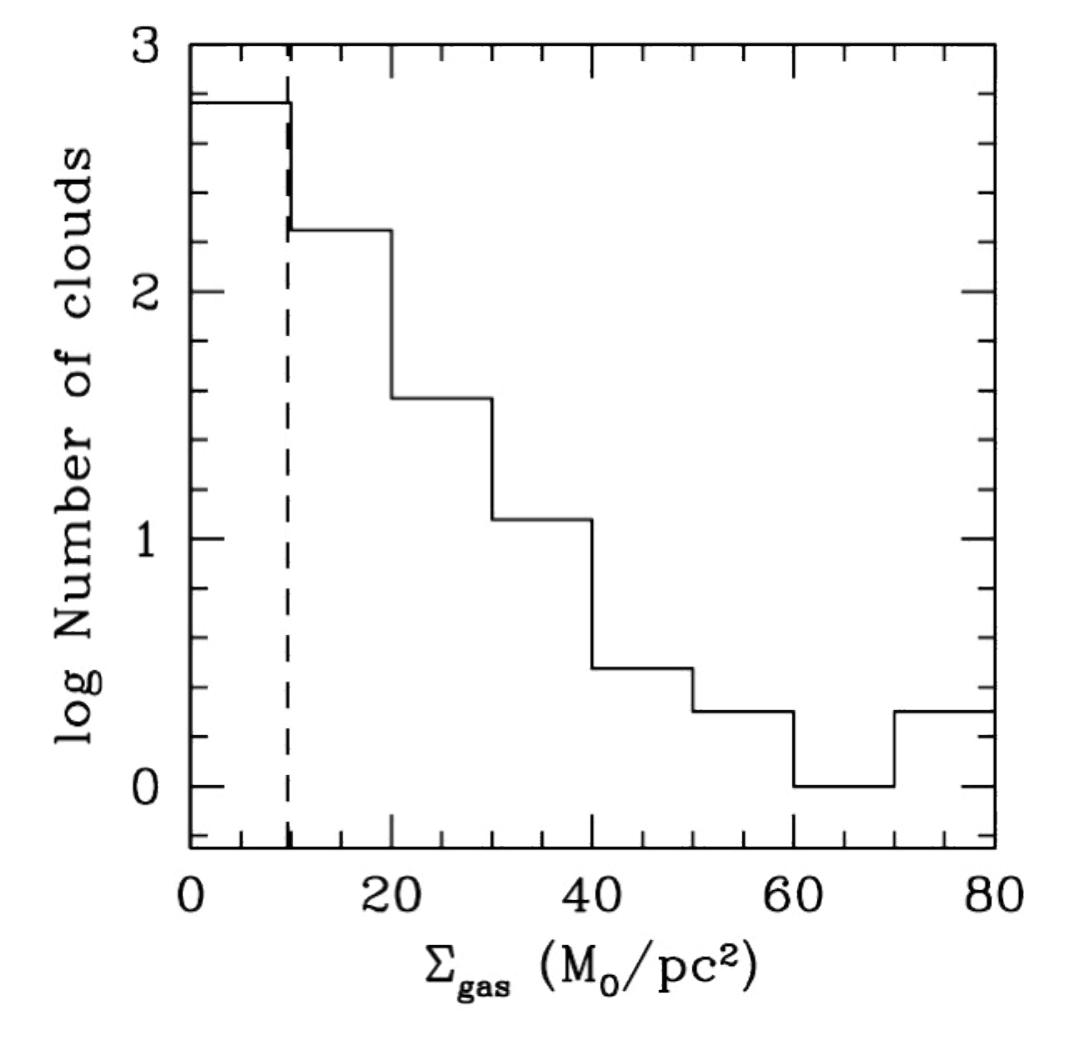}
\caption{Number of clouds from the entire cloud sample with a given gas surface density. The vertical dashed line marks the
value of 9.65 M\solar\ pc$^{-2}$ representing the average over all 814 clouds.
\label{fig-density}}
\end{figure}

In Figure \ref{fig-density} we plot a histogram of the cloud gas surface densities $\Sigma_{\rm gas}$
for all clouds in the sample.
Most clouds have densities less than 10 M\solar\ pc$^{-2}$ and the numbers fall off out to 80 M\solar\ pc$^{-2}$.
The density averaged over all clouds, marked by a vertical dashed line, is 9.65 M\solar\ pc$^{-2}$.
By contrast the average surface density found by \citet{lmc} for LMC clouds was
4.8 M\solar\ pc$^{-2}$, 2.5 M\solar\ pc$^{-2}$, and 1.3 M\solar\ pc$^{-2}$
for their threshold temperatures of 64 K, 32 K, and 16 K, respectively.
\citet{lmc} selected three different brightness temperature thresholds for identifying
pixels belonging to a single cloud in their study: 64 K, 32 K, 16 K.
The lower the brightness temperature, the more clouds grow and merge into one.
Most of our clouds have densities higher than theirs.

\begin{figure}[t!]
\epsscale{1.2}
\plottwo{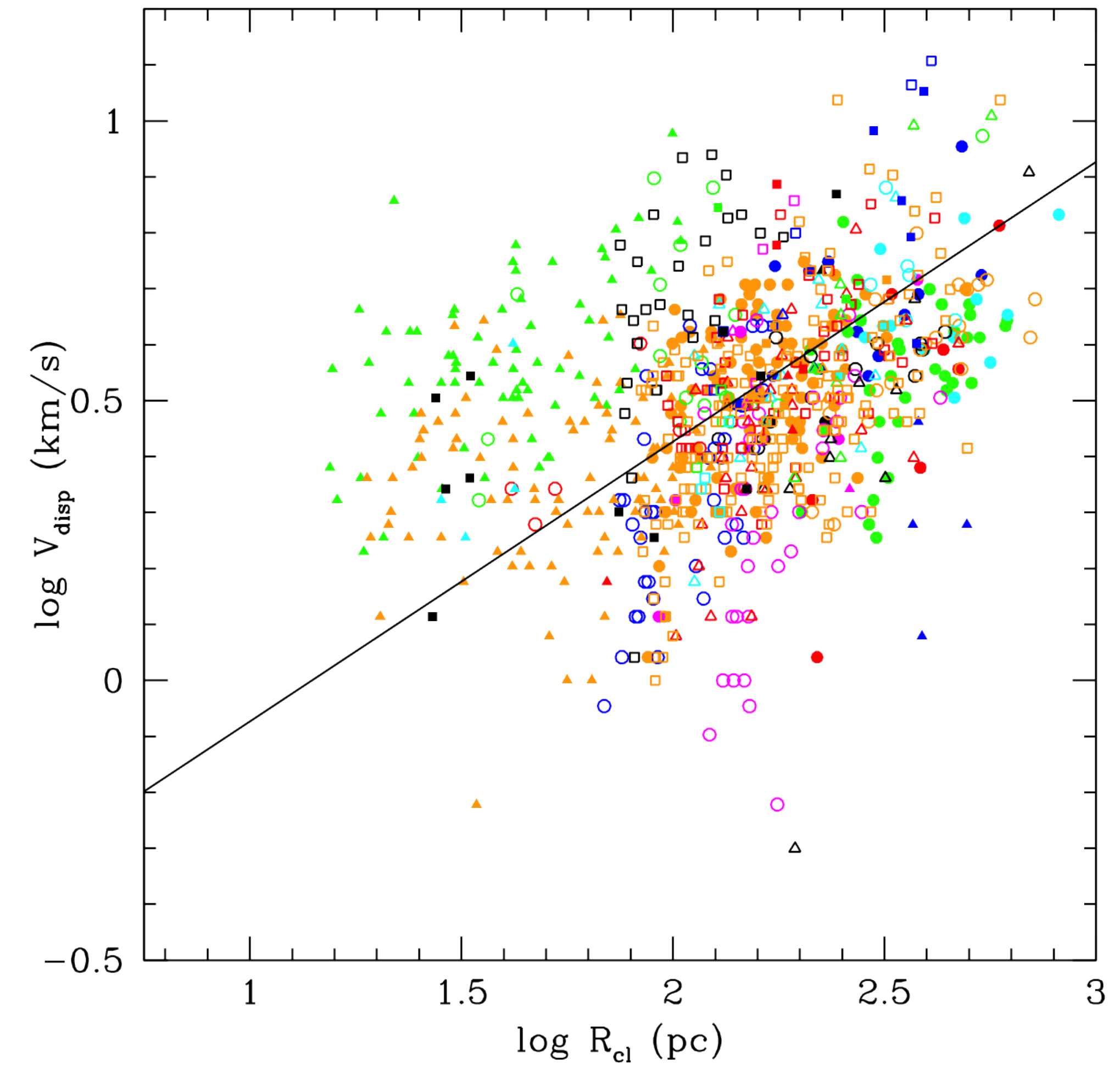}{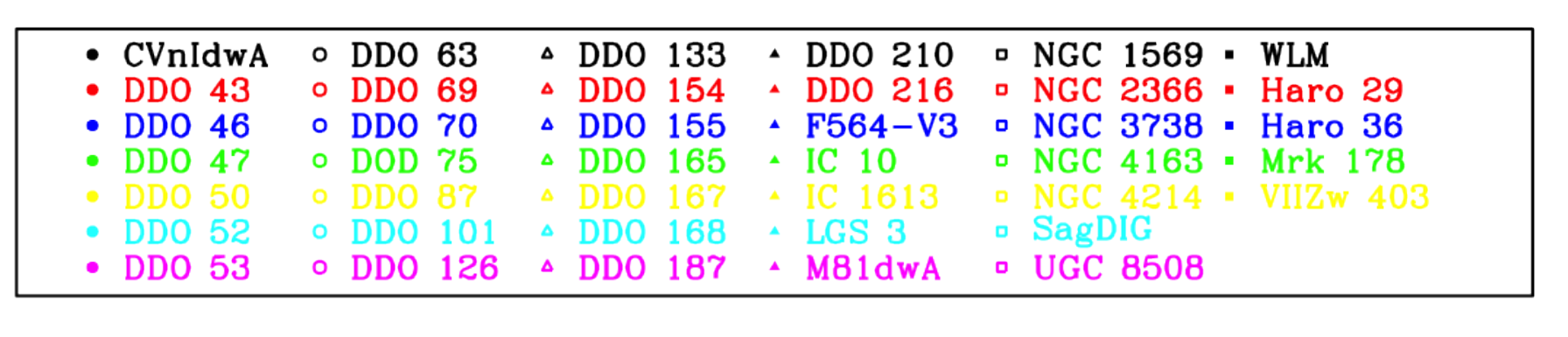}
\caption{Cloud velocity dispersion as a function of cloud radius.
The solid black line has a slope from \citet{lmc} for LMC \HI\ clouds with a constant determined by our cloud sample.
The constant is $0.57\pm0.21$.
\label{fig-sigv}}
\end{figure}

In Figure \ref{fig-sigv} we plot the logarithm of the velocity dispersion $V_{\rm disp}$ as a function of the
logarithm of the size of the cloud \rcl.
There is a very large amount of scatter but generally the larger the cloud, the higher the velocity dispersion.
We plot this in log-log space in order to compare with the results for the LMC by \citet{lmc}.
The solid line in the figure is the LMC relationship plus a constant determined
by our cloud sample:  $\log V_{\rm disp} = 0.5 \times \log R_{\rm cl} - 0.57\pm0.21$.
Our clouds follow this trend but with lots of scatter. Our sizes cover the same range as those for the LMC
clouds that were identified with a threshold of $T_B=32$.

\begin{figure}[t!]
\epsscale{1.2}
\plottwo{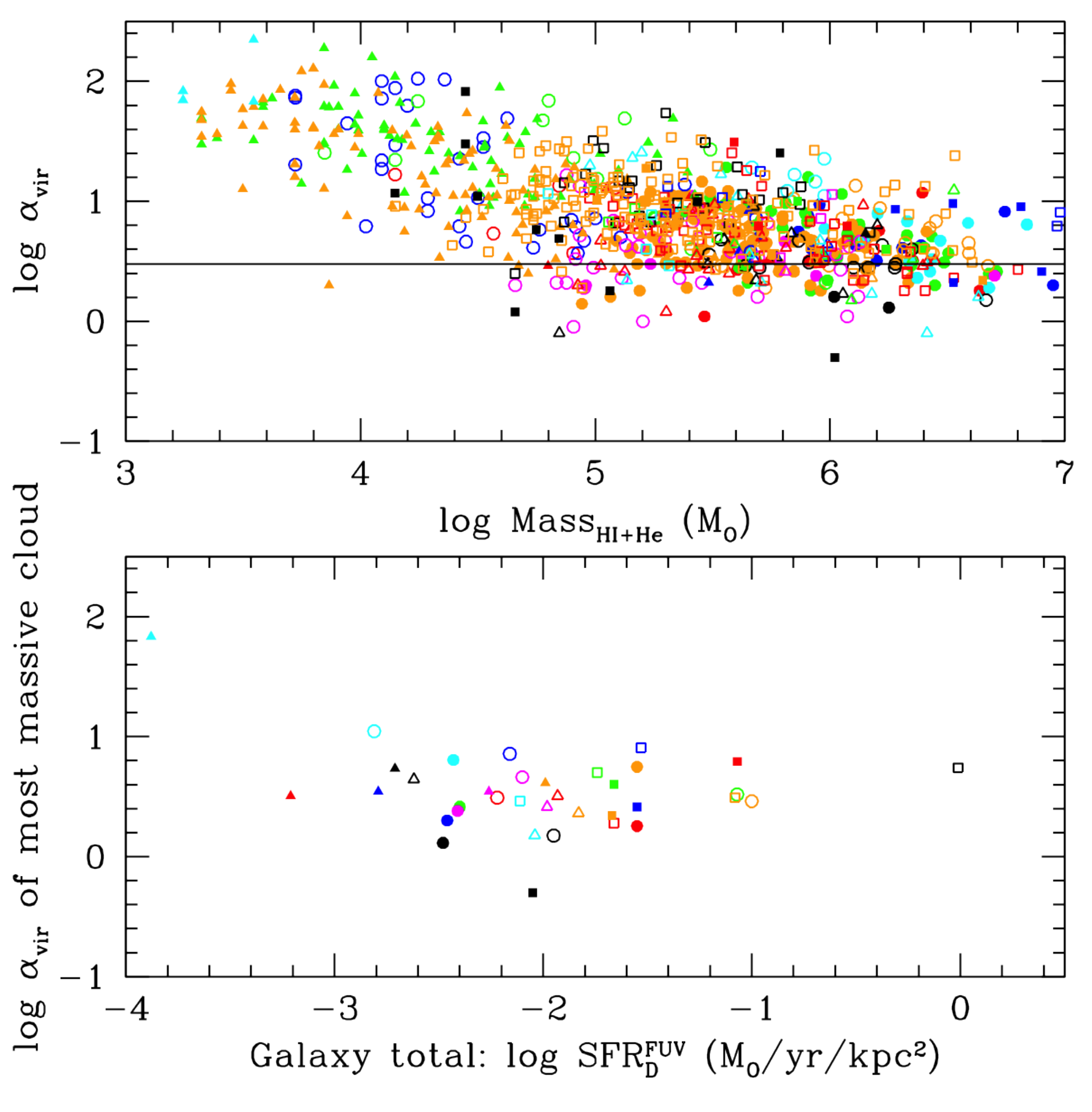}{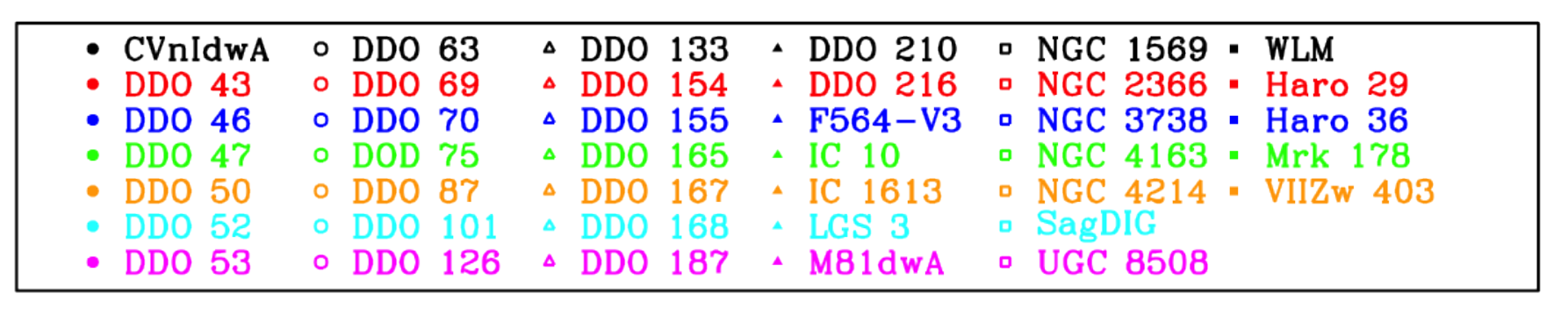}
\caption{Top: \alphavir\ for each cloud plotted against cloud mass.
The solid line marks the limit for most of the clouds identified by \citet{lmc} in the LMC.
Bottom: \alphavir\ of the most massive cloud in each galaxy plotted against the integrated SFR of the galaxy.
\label{fig-alpha}}
\end{figure}

In the top panel of Figure \ref{fig-alpha} we plot \alphavir\  against cloud gas mass for all of the clouds.
The solid horizontal line marks the lower limit for most of the clouds from the \citet{lmc} study of clouds in the LMC.
Our clouds have values like those of the LMC clouds but extend the range to higher \alphavir.
We see a trend of lower \alphavir\ for higher cloud mass.
In the bottom panel of Figure \ref{fig-alpha} we plot the \alphavir\ value of the most massive cloud in each
galaxy against the galaxy integrated star formation rate (SFR). We do not see any trend; there is not much of a range in
\alphavir\ for a large range of SFR.
36\% of the clouds are essentially non-self-gravitating from \HI\
alone, with a virial parameter that exceeds $\alpha_{vir}\sim10$,
and 5\% have $\alpha_{vir}\le 2$.

\subsection{Integrated Properties}

Table \ref{tab-global} gives the number of clouds identified in each galaxy, the total mass in those clouds,
the ratio of cloud mass to total \HI\ mass of the galaxy,
the mass of the third most massive cloud in galaxies with at least three clouds, the fraction of clouds
beyond the break radius $R_{\rm Br}$, the fraction beyond one disk scale length $R_{\rm D}$,
the fraction beyond the extent of \ha\ emission \rha, and the fraction beyond the furthest FUV knot \rfuv.
The break radius is the radius at which the slope of the $V$-band surface brightness profile of the galaxy changes sharply,
usually down-bending \citep{breaks}. Only two of our galaxies do not show such a break in the outer disk: DDO 47 and DDO 210.

\begin{figure}[t!]
\epsscale{0.8}
\plotone{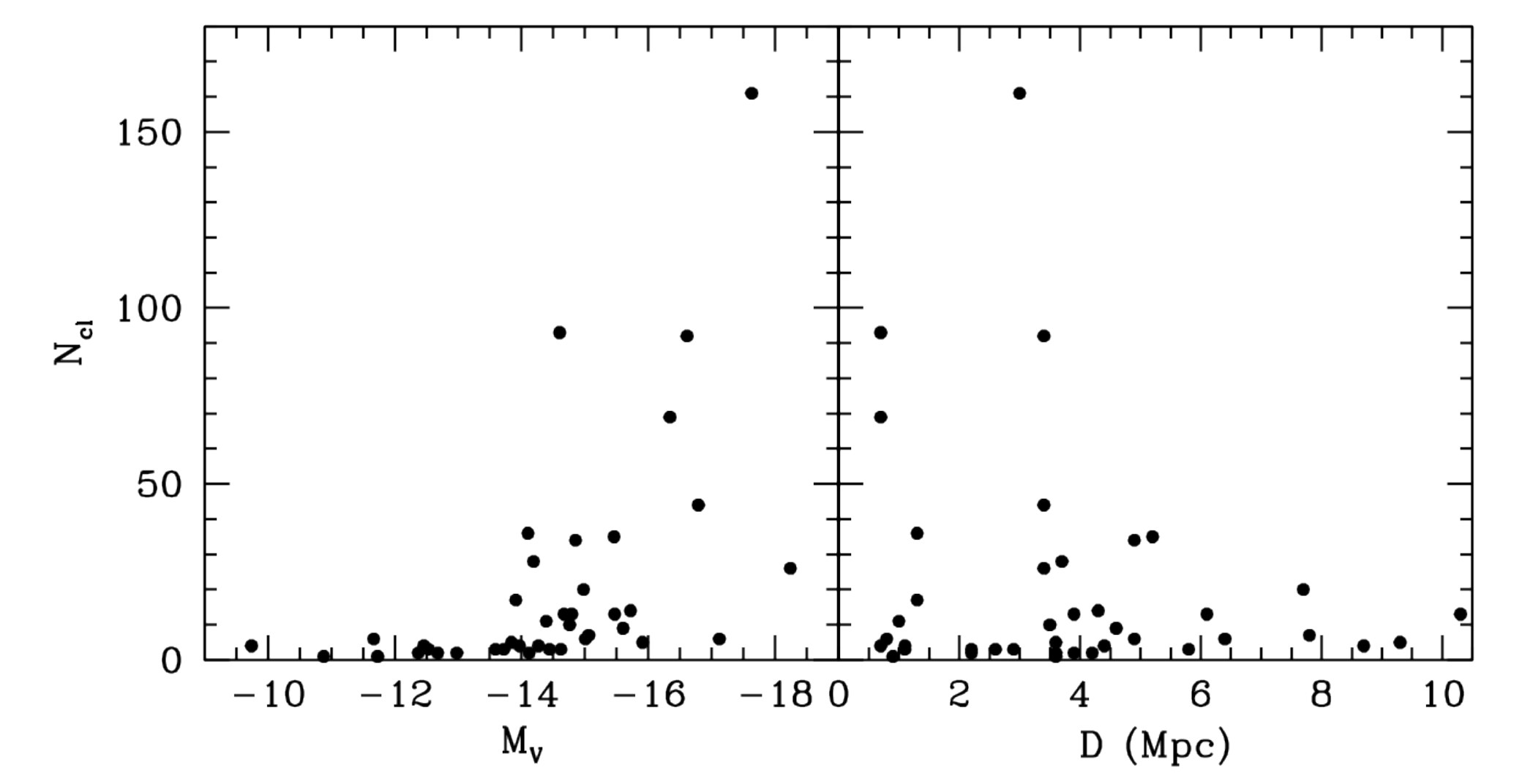}
\vskip -0.1truein
\caption{Number of clouds found in each galaxy versus integrated $V$ magnitude of the host galaxy (left) and
versus the distance to the galaxy (right).
\label{fig-nclvsmv}}
\end{figure}

The number of clouds $N_{\rm cl}$ in a galaxy varies from one cloud in DDO 210 and M81dwA to 161 clouds  in NGC 4214.
The number of clouds is plotted against the integrated absolute $V$ magnitude of the galaxy in Figure \ref{fig-nclvsmv}.
We see that galaxies with M$_V>-14$ have less than about 10 clouds. For galaxies brighter than this, the number of clouds
ranges from a few to hundreds. We also plot $N_{\rm cl}$ against the distance of the galaxy in Figure \ref{fig-nclvsmv}
in order to see if spatial resolution is a factor in how many clouds are identified. We find that the five galaxies with
more than 50 clouds are located within 4 Mpc.

\begin{figure}[h!]
\epsscale{0.7}
\vskip -0.3truein
\plotone{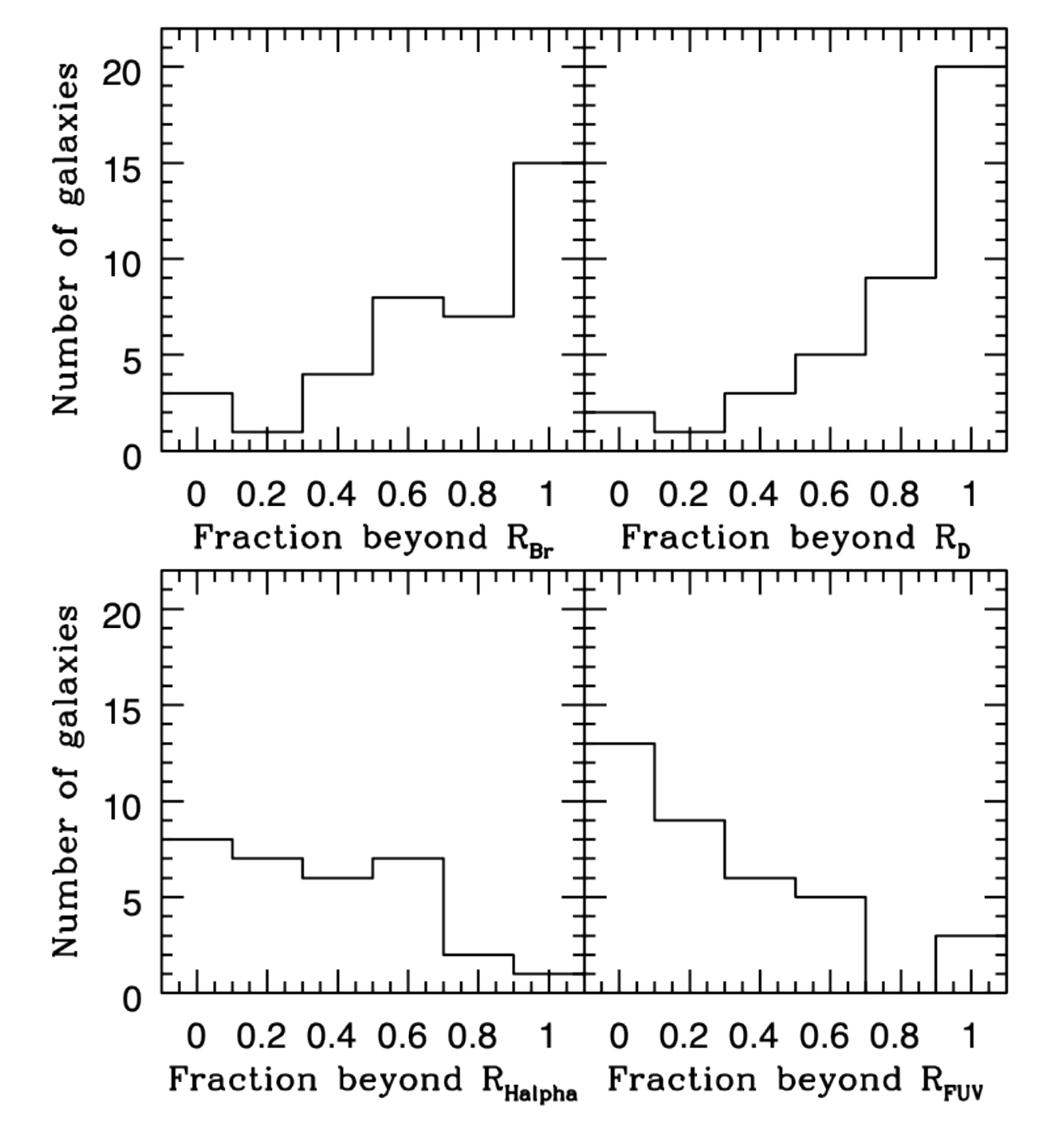}
\vskip -0.3truein
\caption{Number of galaxies with the given fraction of clouds found beyond the galactic break radius \rbr\ (upper left),
beyond one disk scale length \rd\ (upper right), beyond the radius encircling the \ha\ emission \rha\ (lower left),
and beyond the radius of the furthest FUV knot \rfuv\ (lower right).
\label{fig-fraction}}
\end{figure}

The clouds are found from near the center of the galaxy, such as 0.1\rd\ in DDO 126, out to 11\rd\ in DDO 154,
showing that clouds exist both in the inner and in the far outer disk.
In Figure \ref{fig-fraction} we plot a histogram of the number of galaxies as a function of the fraction of clouds beyond \rbr\ and the fraction beyond \rd.
In many galaxies over half of the clouds are found beyond these two radii that mark the outer disk.

\begin{figure}[t!]
\epsscale{0.7}
\vskip -0.2truein
\plotone{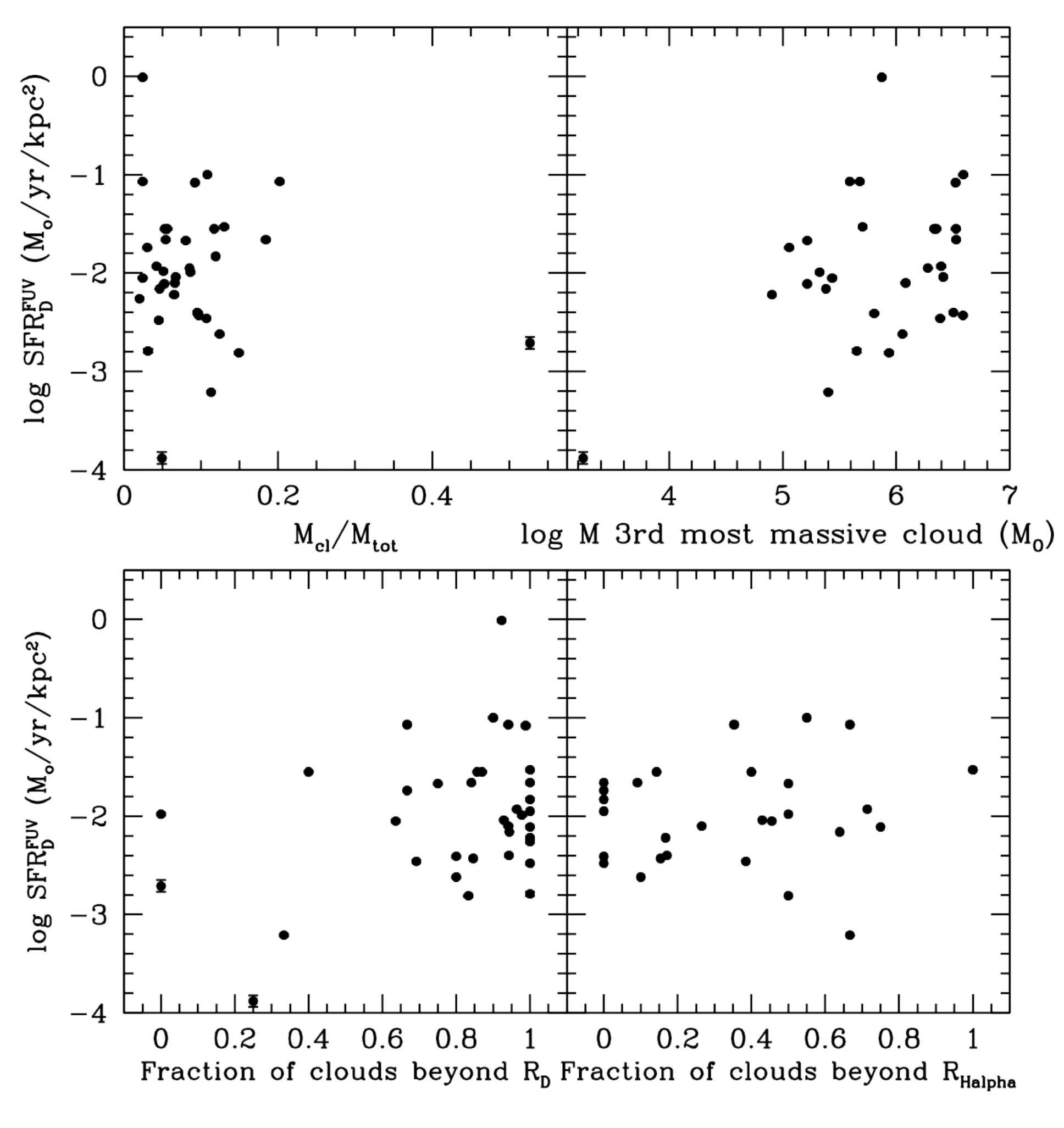}
\vskip -0.2truein
\caption{Galactic SFR surface density plotted against fraction of total gas mass in identified clouds (top left),
the mass of the third largest cloud (top right), the fraction of clouds beyond one disk scale length (bottom left),
and the fraction of clouds beyond \rha\ (bottom right).
\label{fig-sfr}}
\end{figure}

In Figure \ref{fig-sfr} we plot the galactic SFR surface density (integrated
SFR divided by the area in one disk scale length in order to normalize the SFRs)
against various global cloud parameters: fraction of total gas mass in the
identified clouds, the mass of the third largest cloud, and the fraction of
clouds beyond one disk scale length. Aside from a slight trend with the
fraction of clouds beyond $R_{\rm D}$, we do not see any convincing relationship
with SFR surface density.

\subsection{Location of clouds with respect to star forming regions}

In Figure \ref{fig-cvnidwa} we show the location of the clouds on \ha\ images of
those galaxies with \ha\ emission. 
Note that our  \ha\ images do
not cover the entire galaxy for DDO 50, IC 10, IC 1613, and NGC 4214.
Most of the \HI\ clouds in the central regions
are associated with 
or next to \HII\ regions. Of our 40
galaxies, 29 have \HII\ region catalogues from \citet{hiilum}. There are 
349 \HI\ clouds in those galaxies, and of those, 223 clouds, or 64\%, are not
adjacent to or overlapping with an \HII\ region.
Even galaxies without any \HII\ regions (LGS 3, M81dwA, DDO 210), nevertheless, have \HI\ clouds.

In Figure \ref{fig-fraction} the lower left panel shows the number of
galaxies with a given fraction of clouds beyond the extent of detected \ha\
emission \rha\ \citep{sfr}.
Most galaxies have some clouds beyond \rha.
In Figure \ref{fig-sfr} the bottom right panel shows the fraction of clouds beyond \rha\ as a function of the
galaxy-wide SFR surface density of the galaxy. There is no relationship.

\section{Dark Molecular Hydrogen}\label{dark}

Although only a small fraction, $\sim2-53$\%, of \HI\ in these galaxies has been
catalogued here in the form of discrete clouds (cf.\ Fig.\ 13 top left), there is
still an expectation that star formation in the form of \HII\ regions or far-ultraviolet (FUV) should
correlate with clouds of some type, and most likely with dense, self-gravitating
clouds, as observed nearly everywhere in the Milky Way and other nearby spiral
galaxies \citep{mckee07}.  These clouds tend to be CO-emitting molecular clouds
in more metal-rich galaxies like the Milky Way, so it is a reasonable assumption
that CO-weak molecular clouds are associated with star formation in the
low-metallicity dwarf irregulars studied here
\citep[e.g.,][]{leroy11,bolatto13,rubio15,shi15,review19}. Such molecular clouds would not
necessarily show up as discrete \HI\ clouds if their atomic hydrogen is in an
envelope that blends with the \HI\ emission from the diffuse intercloud medium.
This is essentially the model of \cite{shaya87}, namely \HI\ occurs primarily in
the optically thin regions of the interstellar medium and the optically thick gas
is molecular whether or not it shows up in CO.

There are three indications that star formation in our survey is associated with
dark molecular gas. First, a relatively high fraction of the identified \HI\ clouds
are outside the galactic radii where there are clear indications of star
formation. This was shown in Figure \ref{fig-fraction} and by a slight upward
trend in the lower left panel of Figure \ref{fig-sfr}, where higher star
formation rate surface densities have proportionally more of their \HI\ clouds in the outer
regions. Presumably, these high $\Sigma_{\rm SFR}$ regions are rich in molecular
gas. A similar conclusion was made by \cite{leroy08} and others
\citep[see review by][]{review19}.

\begin{figure}[t!]
\epsscale{0.5}
\vskip -0.2truein
\plotone{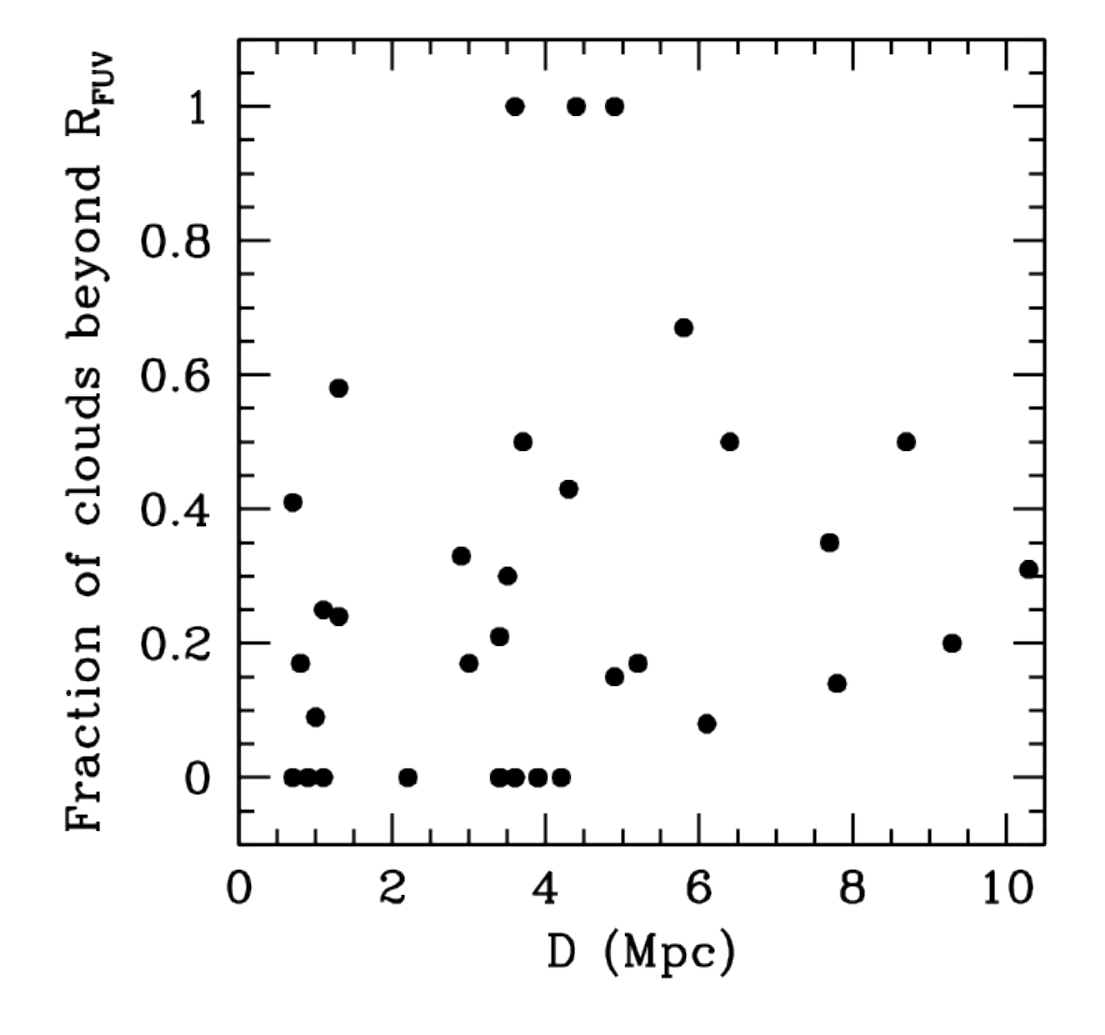}
\vskip -0.2truein
\caption{Fraction of clouds beyond \rfuv\ as a function of distance of the galaxy.
\label{fig-rfuvdist}}
\end{figure}

This point is illustrated further by Figure \ref{fig-fraction} (lower right
panel), which shows a histogram of the number of galaxies with certain fractions
of their \HI\ clouds beyond the edge of the FUV disk. This figure complements
that on the lower left side of Figure \ref{fig-fraction}, which is for
H$\alpha$. (\rfuv\ and \rha\ are given in Table \ref{tab-gal}). In Figure
\ref{fig-fraction}, 22\% of the galaxies have less than half of their
\HI\ clouds in the FUV disk where star formation occurs (i.e., 8/36 of the
galaxies have a value on the abscissa that is greater than 0.5). Similarly,
32\% (10/31) of the galaxies have more than 50\% of their \HI\ clouds
beyond \rha. In these galaxies, most \HI\ clouds are unrelated to star formation
and most star formation must be occurring in clouds that do not show up as
discrete \HI\ objects with our sensitivity and resolution. To examine the
effect that resolution has on this result, we plot the fraction of clouds beyond
\rfuv\ against distance to the galaxy in Figure \ref{fig-rfuvdist}. There we see
that the galaxies within 4 Mpc, which are also the galaxies with highest spatial
resolution for a given cloud size, cover the same range in fraction of clouds
beyond \rfuv\ as the more distant galaxies. The difference is that galaxies with
no clouds beyond \rfuv\ are all within 4 Mpc. We conclude that resolution caused
by distance to the galaxy is not generally skewing our conclusion that a large
fraction of the galaxies in our sample have \HI\ clouds unrelated to star
formation.

\begin{figure}[t!]
\epsscale{0.6}
\vskip -0.1truein
\plotone{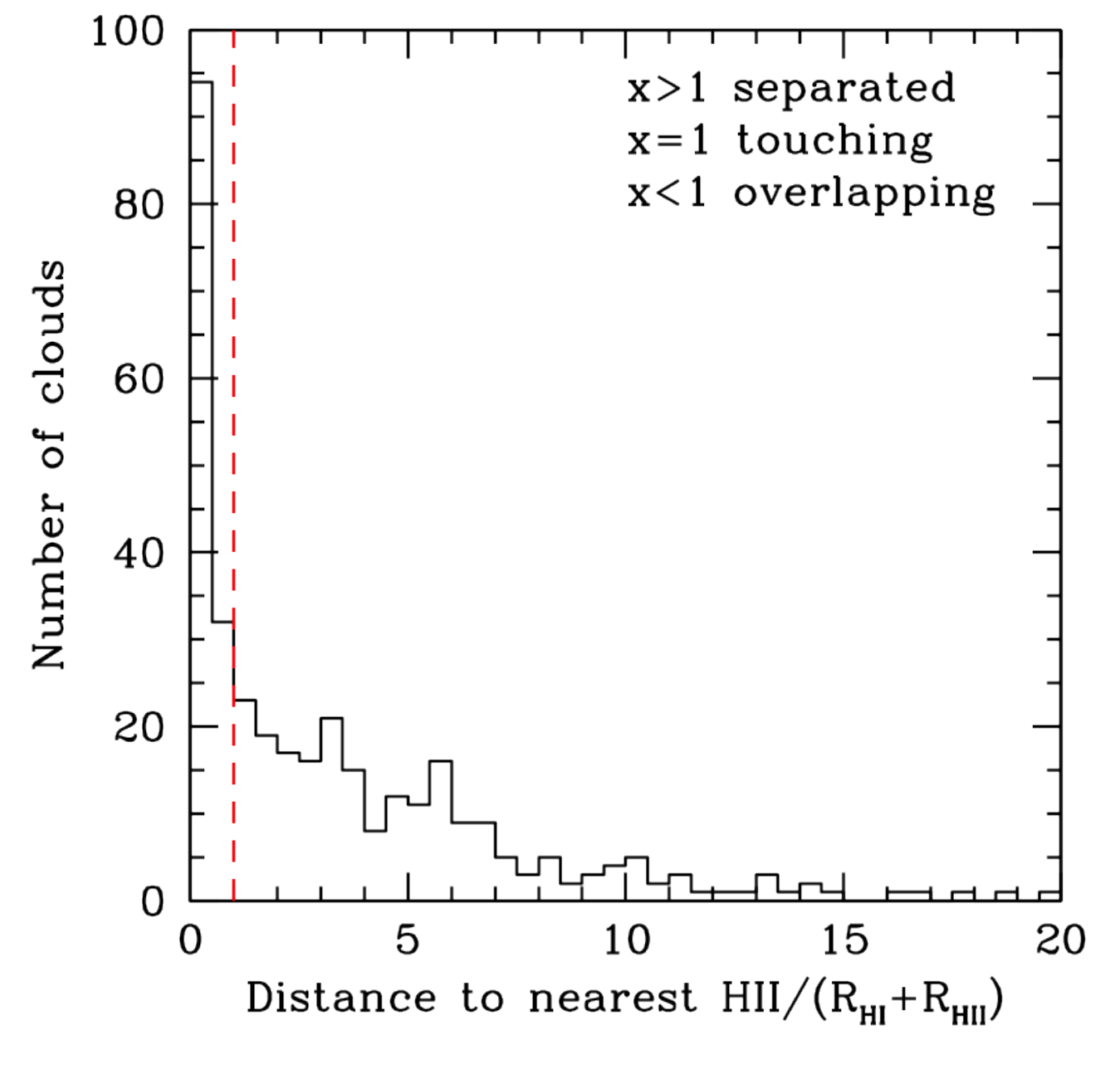}
\vskip -0.3truein
\caption{Histogram of the number of catalogued \HI\ clouds at various distances from their
nearest \HII\ regions, normalized to the sum of the cloud and \HII\ region radii.
Only galaxies with \HII\ region catalogs \citep{hiilum} and \ha\ images that cover the entire
galaxy (omitted DDO 50,IC 10, IC 1613, NGC 4214) were included, for a total of 349 \HI\ clouds.
Approximately
64\% of the clouds are separated from \HII\ regions (abscissa greater than 1), suggesting that
the \HII\ regions are associated with star formation in clouds that are not discrete \HI\ objects.
\label{hist_sep_hi_hii}}
\end{figure}

Secondly, most \HII\ regions lie outside the edges of giant \HI\ clouds or far
from these clouds. Figure \ref{hist_sep_hi_hii} shows a histogram of the number
of \HI\ clouds at certain distances from the nearest \HII\ region, normalized to
the sum of the cloud and \HII\ region radii. When this normalized distance
exceeds unity, which it does for $\sim64$\% of the clouds, the nearest
\HII\ region is associated with star formation that has either destroyed or
pushed away the adjacent regions of its primordial cloud, or has its stars
forming in dark gas of some type. The first possibility seems unlikely for \HII\
regions more remote than a few times their summed sizes because there is little
clearing or destructive force that young stars can exert on a cloud this far from
the \HII\ region, which defines the zone of greatest interaction. Also, the \HII\
regions probably span a range of ages from zero to some 10 Myr, and the youngest
ones that are forming stars now should still have gas nearby
\citep[e.g.,][]{kruijssen18}. Most likely, the \HII\ regions have dark molecular
gas or unresolved \HI\ still fairly close to them.

\begin{figure}[t!]
\epsscale{1.0}
\vskip -0.2truein
\plotone{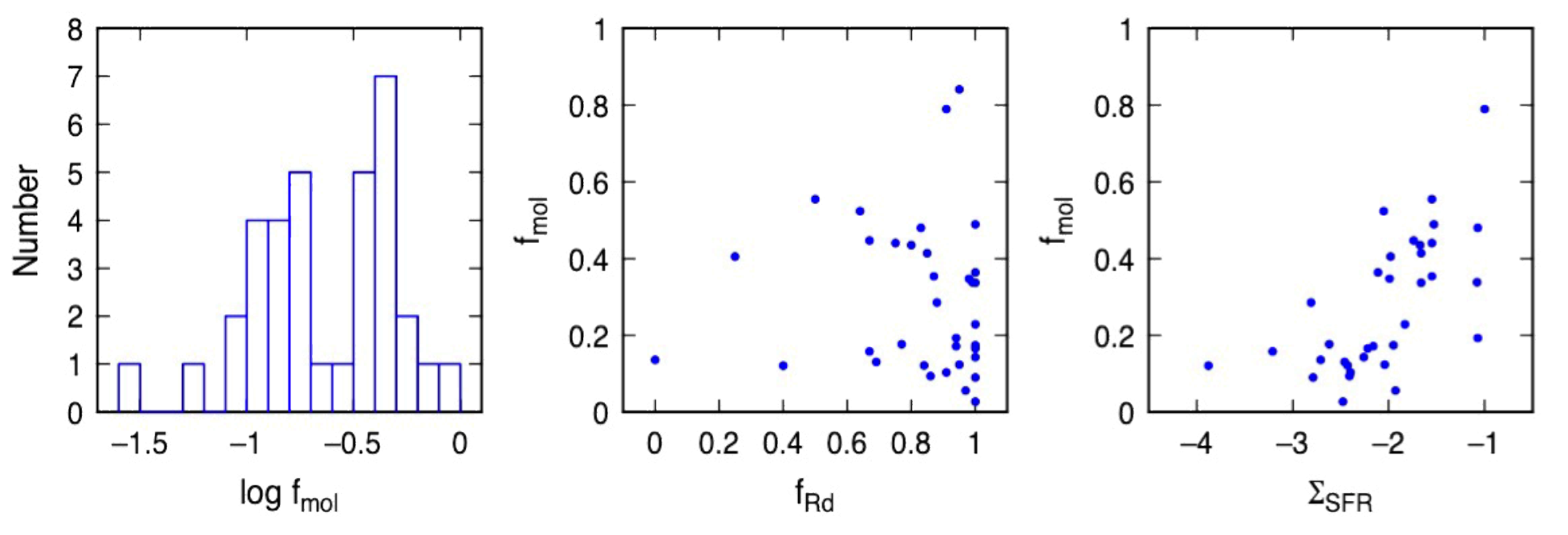}
\vskip -0.1truein
\caption{The inferred total molecular fraction in our galaxies,
calculated from the total star formation rate, the
typical molecular consumption time and the total \HI\ mass, is shown
as a distribution function (left) and in comparison to the fraction of the \HI\ clouds
outside the disk scale length (middle) and the star formation rate surface
density (right). The inferred molecular fraction is typically larger than 10\%, suggesting
a large proportion of unseen molecules in the interstellar media of these dwarf
galaxies, and it is larger for galaxies with most of their cataloged clouds in the outer regions
(i.e., large $f_{\rm Rd}$) and for higher $\Sigma_{\rm SFR}$.
\label{HIcloud_1}}
\end{figure}

Third, the inferred molecular content of these galaxies, determined from the
product of the star formation rate and the typical molecular consumption time of
$t_{\rm consumption}=2\times10^9$ yr \citep{bigiel08}, correlates inversely with
the fraction of \HI\ clouds outside the disk scale length. The total molecular mass
is calculated from the equation
\begin{equation}
M_{\rm mol}\approx\Sigma_{\rm SFR_{\rm D}}^{\rm FUV}\pi R_{\rm D}^2 t_{\rm consumption}
\label{mmol}
\end{equation}
and the molecular fraction is calculated from
\begin{equation}
f_{\rm mol}={{M_{\rm mol}}\over{{M_{\rm mol}+M_{\rm HI\;tot}}}},
\end{equation}
using quantities from Table \ref{tab-gal}. The distribution function of this
molecular fraction is shown on the left in Figure \ref{HIcloud_1}. Typically more
than 10\% of the total gas is molecular according to this calculation; the
average is $\log f_{\rm mol}=-0.64$, or $f_{\rm mol}\sim23$\%.  The
molecular fraction is also shown versus the fraction of catalogued clouds outside
one disk scale length in the middle panel and versus the star formation 
rate surface density on the right. The middle panel shows that the galaxies
with the highest molecular fractions also have the highest fraction of their
clouds in the outer regions, which implies that the missing molecules are in the
inner regions. The right-hand panel shows an increasing trend that follows mostly
from the assumption that the molecular mass is proportional to the star formation
rate, but it also contains the \HI\ mass inside of $f_{\rm mol}$, and \HI\ is not
part of that assumption.  The implication is that regions with higher
$\Sigma_{\rm SFR}$ have a greater proportion of unobserved gas, most likely
molecular, if they satisfy the Kennicutt-Schmidt relation \citep{bigiel08}.
Otherwise they would lie high above this relation with star formation occurring
in relatively little gas.


The distributions and correlations in Figures \ref{fig-fraction}, \ref{hist_sep_hi_hii}, and \ref{HIcloud_1}
support the idea that a significant fraction of gas in the inner regions of dwarf
irregulars is in molecular form as H$_2$, even though it is not yet visible in CO
or other molecular tracers nor clearly visible in dust emission because of
low metallicity and temperature.  Most likely, stars are forming in a
combination of CO-dark H$_2$ gas and atomic gas that is too cold and optically
thick to stand out in our emission line maps.

The additional gas mass inferred from equation (\ref{mmol}) makes the
interstellar media of these dIrrs look more unstable to self-gravitational
processes than previously thought. In \cite{aomawa98}, we calculated for many of
the same galaxies that the area-average ratio of the gas surface density,
$\Sigma_{\rm g}$, to the critical value from \cite{toomre64}, $\Sigma_{\rm c}$,
was 0.53 (see Table 3 in that paper), compared to 0.7 at the H$\alpha$-emitting
edges of spiral galaxies in \cite{kennicutt89}.  Figure \ref{HIcloud_1} now
suggests that the average gas mass inside $R_{\rm D}$ is higher by a factor
$1/(1-<f_{\rm mol}>)=1/(1-0.23)=1.3$ when invisible H$_2$ is included. This
factor of 1.3 brings our previous value of $\Sigma_{\rm g}/\Sigma_{\rm c}$ up to
0.69, close to the Kennicutt value. Thus it seems possible that gravitational
instabilities initiate star formation in some dIrrs like they do in spirals, and that
dIrrs do this in a diffuse atomic and molecular medium that is currently
invisible in CO emission. This is consistent with our previous suggestion
\citep{elmegreen15} that the average midplane density of gas in dIrrs is higher
than inferred from \HI\ alone as a result of an invisible molecular diffuse medium.

\section{Summary}

The properties of 814 \HI\ clouds in 40 nearby dIrr galaxies are determined from
automated cloud-finding software and compared with the distributions of \HII\
regions and FUV emission from young stars.  The internal cloud properties suggest
a surface density spanning a relatively small range with a cloud average value,
including He, equal to $\sim9.65\;M_\odot$ pc$^{-2}$. Cloud masses range from
$\sim10^3\;M_\odot$ to $10^7\;M_\odot$ with the larger clouds in more massive
galaxies. There is a weak square-root relation between internal cloud velocity
dispersion and size, similar to that for molecular clouds, although the
dispersions for these \HI\ clouds are large enough to make most of them only
weakly self-gravitating or diffuse.

The most interesting result is that the \HI\ clouds tend to be outside the main
regions of star formation, as shown by their prominence in the outer disks or by
their typically large separations from the nearest \HII\ regions.  This suggests
that star formation is occurring in different clouds that are not visible in our
\HI\ emission line survey. Considering the usual correlation between molecular
hydrogen as traced by CO emission and the star formation rate, we can infer the
mass of missing molecular gas in these dIrrs. This missing H$_2$ increases the
gas mass in the inner disks by a galaxy average value of 23\%, and
makes the gas there about as unstable gravitationally as the gas in spiral
galaxies where CO is observed directly.  We conclude that the inner, star-forming
disks of dIrr galaxies are rich in molecular and opaque atomic gas that does not
typically appear in \HI\ surveys, and that star formation proceeds by fairly
normal processes in the dense molecular parts of these disks.


\acknowledgments We are grateful to E. Rosolowsky and A.K. Leroy for making their
CLOUDPROPS code available to all. CLB appreciates funding from the National
Science Foundation grant AST-1461200 to Northern Arizona University for Research
Experiences for Undergraduates summer internships and Dr.\ David Trilling for
running the NAU REU program in 2017.  \ Facilities: \facility{VLA}.

\end{document}